\documentclass[traditabstract]{aa} % for the abstract without structuration 
\usepackage{graphicx}
\usepackage[export]{adjustbox}
\usepackage{amsmath}
\usepackage{txfonts}
\usepackage{natbib}
\bibpunct{(}{)}{;}{a}{}{,}
\usepackage{float}
\usepackage{calc}
\usepackage{textcomp}
\usepackage{placeins}
\usepackage{color}
\usepackage{rotating} 
\usepackage{lscape}
\usepackage{url}
\usepackage{hyperref}
\usepackage{multicol}
\usepackage{subfig}
\usepackage[all]{draftcopy}
\usepackage{longtable}
\usepackage{array}
\usepackage{threeparttable}
\usepackage{lscape}
\DeclareTextSymbol{\degre}{OT1}{23}
\newcounter{savedfootnote}

\begin{document}
   \title{Morphological Classification of Local Luminous Infrared Galaxies}

\author{ A. Psychogyios\inst{\ref{inst1}}  \and  V. Charmandaris\inst{\ref{inst1},\ref{inst2}}\and T. Diaz- Santos\inst{\ref{inst3}}\and L. Armus\inst{\ref{inst4}}\and S. Haan\inst{\ref{inst5}} \and J. Howell\inst{\ref{inst6}} \and E. Le Floc{'}h \inst{\ref{inst7}} \and S. M. Petty\inst{\ref{inst8}} \and A. S. Evans\inst{\ref{inst9},\ref{inst10}} }

	\institute{Department of Physics, University of Crete, GR-71003 Heraklion, Greece\\\email{alpsych@physics.uoc.gr}\label{inst1}
	\and IAASARS, National Observatory of Athens, GR-15236, Penteli, Greece \label{inst2} \and N\'ucleo de Astronom\'ia de la Facultad de Ingenier\'ia, Universidad Diego Portales, Av. Ej\'ercito Libertador 441, Santiago, Chile \label{inst3} \and  Spitzer Science Center, Calfornia Institute of Technology, MS 220-6, Pasadena, CA 91125 \label{inst4}\and CSIRO Astronomy and Space Science, ATNF, PO Box 76, Epping 1710, Australia \label{inst5} \and Infrared Processing and Analysis Center, California Institute of Technology, MS 100-22, Pasadena, CA 91125, USA \label{inst6} \and CEA Saclay, DSM/Irfu/Service d{'}Astrophysique, Orme des Merisiers, F-91191 Gif-sur-Yvette Cedex, France \label{inst7} \and Department of Physics, Virginia Tech, Blacksburg, VA 24061, USA \label{inst8} \and Department of Astronomy, University of Virginia, Charlottesville, VA 22904, USA \label{inst9} \and  National Radio Astronomy Observatory, Charlottesville, VA 22903, USA \label{inst10}}

   \date{Received  / Accepted}

%=================================================================================
%================================   ABSTRACT  ======================================
%=================================================================================
 
  \abstract{
We present an analysis of the morphological classification of 89 luminous infrared galaxies (LIRGs) from the Great Observatories All-sky LIRG Survey (GOALS) sample using non-parametric coefficients and compare their morphology as a function of wavelength. We rely on images obtained in the optical (B- and I-band) as well as in the infrared (H-band and 5.8$\mu$m). Our classification is based on the calculation of $Gini$ and the second order of light ($M_{20}$) non-parametric coefficients which we explore as a function of stellar mass ($M_\star$), infrared luminosity ($L_{IR}$) and star formation rate (SFR). We investigate the relation between $M_{20}$, the specific SFR (sSFR) and the dust temperature ($T_{dust}$) in our galaxy sample. We find that $M_{20}$ is a better morphological tracer than $Gini$, as it allows to distinguish systems formed by double systems from isolated and post-merger LIRGs. The effectiveness of $M_{20}$ as a morphological tracer increases with increasing wavelength, from B- to H-band. In fact, the multi-wavelength analysis allows us to identify a region in the $Gini$-$M_{20}$ parameter space where ongoing mergers reside, regardless of the band used to calculate the coefficients. In particular when measured in the H-band, this region can be used to identify ongoing mergers, with a minimal contamination from LIRGs in other stages. We also find that while the sSFR is positively correlated with $M_{20}$ when measured in the mid-infrared, i.e. star-bursting galaxies show more compact emission, it is anti-correlated with the B-band based $M_{20}$. We interpret this as the spatial decoupling between obscured and un-obscured star formation, whereby the ultraviolet/optical size of a LIRGs experience an intense dust enshrouded central starburst is larger than in the one in the mid-infrared since the contrast between the nuclear to the extended disk emission is smaller in the mid-infrared. This has important implications for high redshift surveys of dusty sources, where sizes of galaxies are routinely measured in the rest-frame ultraviolet.
}
   
   \authorrunning{Psychogyios et al. 2015 }
   \titlerunning{Morphological Classification of Local Luminous Infrared Galaxies}

  \maketitle

%=================================================================================
%================================  INTRODUCTION   =================================
%=================================================================================
\section{\label{intro}Introduction}

One of the outstanding questions in extragalactic astronomy is how matter in the universe assembled into the structures we see today. An approach to tackle this question is to study the formation and evolution of galaxies, since they are luminous beacons of the baryon content of the Universe. Galaxy morphology, which traces how the electromagnetic emission of the various physical processes is distributed across a galaxy, can be used to study how galaxies evolve and which is the dominant mechanism shaping this evolution. Fundamental properties of galaxies, such as their mass, baryonic content, star formation history, interaction state and environment are intimately connected  with galaxy morphology \citetext{\citealp{Dressler80, RobertsHaynes94, Kennicutt98, Strateva01, Wuyts11}}.
Morphological studies can be used to constrain the theoretical models of galaxy evolution. \\
\indent Traditionally the \textquoteleft Hubble Tuning Fork\textquoteright  \citep{Hubble26} provides significant information about the morphology of bright and massive galaxies in the local universe and is closely correlated with galaxy physical properties as stellar mass ($M_\star$), color, star formation rate (SFR) and relative dominance of a central bulge \citep{RobertsHaynes94}. However, using it to classify galaxies at z > 1 is challenging due to limitations in angular resolution and progressive decreasing of the signal to noise as a result of both the surface brightness dimming and the sampling of shorter wavelengths at a given observed band-pass \citep{vandenBergh96, Dickinson00}. A number of optical studies observed systems at z > 2 with Hubble Space Telescope (HST), by sampling their rest-frame ultraviolet (UV) emission, revealed that many high-redshift galaxies exhibit irregular shapes and do not follow the typical Hubble types \citetext{\citealp{Lotz04, Papovich05, Lotz06, Conselice08}}.  As a result, other methods using parametric coefficients, such as the Sersic index ($n$) \citep{Sersic68} or non-parametric coefficients, like $Gini$ and $M_{20}$ \citetext{\citealp{Abraham03, Lotz04}} have been developed in order to quantify the morphology of a galaxy.\\
\indent At z$\sim$1, corresponding to a \textquoteleft look-back time\textquoteright  of nearly 8 Gyr, luminous infrared galaxies (LIRGs) begin to dominate the IR background and the star formation rate density (SFRD - \citep{Magnelli13}). These galaxies emit a higher fraction of energy in the infrared (IR) spectrum ($\sim$5-500 $\mu$m) than at all other wavelengths combined. A LIRG by definition emits more than $10^{11}$$L_{\sun}$ in the IR (8-1000$\mu$m) part of the electromagnetic spectrum, while, a more luminous system, emitting more than $10^{12}L_{\sun}$, is called ultraluminous infrared galaxy (ULIRG) \citep{SanMir96}. The power source of most local (U)LIRGs is a mixture of accretion on to an Active Galactic Nucleus (AGN) and a circumnuclear starburst, both of which are fueled by large quantities of high density molecular gas that has been funneled into the merger nucleus. In the process of a violent interaction of two spiral galaxies, hydrogen clouds that were initially distributed throughout the galactic disc could move to the centre forcing the gas to become concentrated. Numerical simulations of colliding galaxies \citep{Barnes92,MihosHernquist96,Hopkins08} showed that the gas and stars react differently during a merger. The gas tends to move out in front of the stars as they orbit the galactic centre. Furthermore, gravitational torques on the gas reduce its angular momentum, causing it to plunge toward the galactic centre. As the two galaxies begin to merge, more angular momentum is lost and the concentrated circumnuclear gas feeds a massive starburst and/or an AGN. \\
\indent Observations with the Infrared Space Observatory (ISO) and \textit{Spitzer} Space Telescope showed that they contribute up to 50\% of the cosmic infrared background, dominating the SFR of the universe at z$\sim$1 \citetext{\citealp{Elbaz02, LeFloch05, Caputi07, Magnelli09}}. Despite the rarity of (U)LIRGs in the local Universe, their study is of paramount importance as they allow exploration of their detailed morphologies that cannot be done (owing to resolution limitations) at higher redshifts where they are order of magnitudes more common \citetext{\citealp{Blain02, Chapman05}}. \\
\indent Furthermore, (U)LIRGs are strongly related with the evolution of massive galaxies. Mergers both major and minor are important in the morphological transformation of galaxies since they transforming spiral disks into red and spheroids, building the high mass end of the stellar mass function, mostly at moderate to low redshifts \citep{Williams11}. In particular, a system of two spiral galaxies that interact dynamically will pass through a violent stage, where the spiral arms and the disc of both galaxies will be destroyed and as consequence of the violent relaxation the population of the stars will relax to an $r^{1/4}$ profile, which is characteristic distribution of an elliptical galaxy \citep[i.e.][]{Hjorth91,Hopkins09}. As a consequence unlike high-z (U)LIRGs which are mainly isolated systems which extended gas rich disks intensely forming stars, local LIRGs exhibit a large range of morphologies, from isolated galaxies to interacting pairs and mergers. \\
\indent The morphological study of LIRGs across a wide wavelength range can provide information on the dynamical history of galaxies \citep[i.e.][]{Lotz04,Petty14}. When combined with spectra or spectral energy distribution (SEDs), the morphologies indicate how galactic environment (or the merger history) has influenced star formation (SF). In particular, UV light originates from the young massive OB stars in star-forming regions while the optical light is mainly emitted by less massive stars. The near infrared (NIR) part of the electromagnetic spectrum unveils the location of the older stellar populations responsible for bulk of the total mass. \\
\indent With the advent of a new generation of telescopes such as JWST \citep{Gardner06}, Euclid \citep{Amiaux12}, LSST \citep{LSST12} and the Dark Energy Survey \citep{Frieman13} huge amounts of data for millions of galaxies will be available at higher redshifts. It will be essential to have robust tools for automatic morphological and structural classifications. In this paper we investigate the connection of the local LIRG morphologies with other physical properties and pave the way for studying their high-z analogues. Our motivation in this work is to refine the morphological method of $Gini$ and $M_{20}$ for galaxies that are dusty and suffer an ongoing merger stage. \\
\indent The paper is organised as follows. In Sect. 2 we describe the data. The morphological diagnostics are described in the Sect. 3. In Sect. 4 we present a summary of the analysis we performed in order to calculate the non-parametric coefficients. We present our results in Sect. 5 and discuss about them in Sect. 6 while we compare our findings with other similar studies. Finally, we summarize our conclusions in Sect.7. In Appendix A we also provide the $Gini$ and M$_{20}$ values of our sample using the full B- and I-band field of view, while in Appendix B we provide a brief discussion based on the values of $M_{50}$ coefficient.

%=================================================================================
%================================   SAMPLE & DATA  =================================
%=================================================================================
\section{Sample and Data}
		
	\subsection{Sample}
The sample upon which we base our analysis consists of 89 LIRGs and ULIRGs from the Great Observatories All-sky LIRG Survey \citep[GOALS,][]{Armus09}. GOALS is a sample of 202 LIRGs selected from the IRAS Bright Galaxy Sample (RBGS; \citet{Sanders03}), and spans a redshift range of 0.009 < $z$ < 0.088. The RBGS contains all the extragalactic objects with $S_{60}$ > 5.24Jy observed by IRAS at galactic latitudes $|b|\degr>5$. The infrared luminosities of our 89 LIRGs lie above the value of  $log(L_{IR}/L_{\sun})>10.44$ , the luminosity at which the local space density of LIRGs exceeds that of optically selected galaxies. These galaxies are the most luminous members of the GOALS sample and they are predominantly mergers and strongly interacting systems. The range of infrared luminosity is 10.44 < $\log( L_{IR}/L_{\sun})$ <12.43, with a median of 11.62. The measurements of $L_{IR}$ were taken from \citet{DiazSantos13}. Our final GOALS sample contains 89 galaxies.
	\subsection{Data}
All galaxies in our sample have imaging obtained with the Advanced Camera for Surveys (ACS), the  Wield Field Camera (WFC), the Near Infrared Camera and Multi-Object Spectrometer (NICMOS) and the Wide Field Camera 3 (WFC3) of the HST.  Additionally, mid-IR (MIR) imaging is available with the Infrared Array Camera (IRAC) of $Spitzer$.
	\subsubsection{ACS HST observations}
Our sample has been imaged with the ACS/WFC using the F435W (B) and F814W (I) broad-band filters (GO program 10592, PI: A. Evans: see Evans et al. 2013). The F435W and F814W observations were performed using a three and two point line dither patterns, respectively. The processed B- and I-band images of our sample have a large 202$''$ $\times$ 202$''$ field of view and was selected to capture the full extent of each interaction in one HST pointing. These filters have pixel scales of 0.05$''$. At the median redshift of our sample (z=0.033) 1$''$ corresponds to $\sim$660pc, hence the ACS FoV covers a projected area of $\sim$ 133 kpc$'$ x 133 kpc$'$.The B band has $\lambda_{cen}$= 4297 $\AA$ and width of 1038 $\AA$. Accordingly, the I band has $\lambda_{cen}$= 8333 $\AA$ and width of 2511 $\AA$. Further details of the observations and data reduction are described in \citet{Kim13}.
	\subsubsection{HST NICMOS observations}
HST images with NICMOS/NIC2 have been obtained using the F160W filter for 80 LIRGs. The data were collected using camera two (NIC2) with a field of view (FoV) of 19.3$''$ $\times$ 19.5$''$, a pixel size of 0.075$''$ and are dithered to yield a total field area of typically 30$''$ $\times$ 25$''$. Note that NICMOS failed during execution of this program, so that not all targets in the complete sample of 89 LIRGs were observed. The remaining 9 have been observed with WFC3, which has a wider FoV of 123$''$ $\times$ 136$''$ and a pixel size of 0.13$''$. We refer the reader to \citet[hereafter H11]{Haan11} for more details on the NICMOS data reduction. 
	\subsubsection{IRAC observations}
IRAC is a four-channel camera  \citep{Fazio04} on board $Spitzer$, that provides 5.2$'$ $\times$ 5.2$'$ images at 3.6, 4.5, 5.8, and 8 $\mu$m. In our analysis, we used the images of IRAC channel 3 at 5.8$\mu$m. The detector has 256 $\times$ 256 pixels, with a pixel size of 1.2$''$, corresponding to 844 pc for the median sample redshift. \\

%=================================================================================
%================================   Morphological Diagnostics  =================================
%=================================================================================
\section{Morphological Diagnostics}
A number of methods are used to describe the morphology of galaxies based on the distribution of their emitted radiation. Visual morphology \citep{Sandage05,Lintott11} is the traditional approach in understanding the structure of galaxies. The major classification system in use today was proposed by \citet{Hubble26}, and updated by \citet{deVaucouleurs59} and \citet{Sandage61,Sandage75}\footnote {A review of galaxy classification can be found in \citet{Buta13}}. On the other hand, computer algorithms have been developed to quantify the morphology of galaxies in a faster and possibly less biased manner \citep{SchutterShamir15}. Additionally to the visual and algorithm way of characterising morphologies, galaxies can be classified with the use of various proxies with either parametric or non-parametric coefficients.

\subsection{Methods using parametric coefficients}
This type of method requires a prescribed analytic function, in order to quantify the morphological type. This analytic function is used to model the projected light distribution and quantify the galaxy morphology with a few parameters. Historically one of the first ways to classify the structure was based on integrated light profiles. With this method a single or multiple components can be used to model the galaxy profile \citep{Blanton03ApJ594,Allen06MNRAS371,
Haussler13}.
\citet{Sersic68} empirically derived the following function to describe the radial surface brightness profile. \\
\begin{equation}
I(r) ~ = ~ I_{\rm e} \exp \left\{ -b_{n}  \left[  \left(  \frac{r}{r_{\rm e}} \right)^{1/n} -1 \right]  \right\} ,
\end{equation}
where $r_{\rm e}$ is the effective or half-light radius, $I_{\rm e}$ is the intensity at the effective radius, $n$ is the Sersic index and $b_{\rm n}$ is a function of $n$ \citep{Graham05}. The term $r_{\rm e}$ relates to the physical size of the galaxy and $n$ gives an indication of the concentration of the light distribution.
The value of $n$ is used to describe galactic structures such as bars n$\sim$0.5, disks n$\sim$1, bulges (1.5 < n < 10 ) and  even the light profile of elliptical galaxies (1.5 < n < 20).\\

\subsection{Methods using non-parametric coefficients}
Non parametric methods do not assume an analytical function for the description of the distribution of light in a galaxy. For that particular reason they can be applied to spiral or elliptical galaxies as well as to disturbed systems displaying features of dynamical interaction such as tidal tales, or bridges. Furthermore, non-parametric coefficients are less affected by limitations of resolution and can be measured out to high redshifts, making them ideal for exploring galaxy evolution accross cosmic time \citep{Conselice14}. 
Since most (U)LIRGs are members of interacting systems and they often exhibit irregular shapes, our analysis is based on non-parametric coefficients. \\
\indent There are many non-parametric coefficients in the literature which can be applied to morphological studies. One of them, the CAS classification system has been used extensively during the last decade. The three coefficients of CAS are the following: The concentration index (C), which was developed by \citet{Abraham94,Abraham96}, measures the ratio of the inner and outer part of the light in a galaxy and it correlates with the bulge to disc (B/D) galaxy ratio. \citet{Schade95} defined the asymmetry (A) coefficient and used it to automatically distinguish spirals from ellipticals and galaxies with irregular shapes. \citet{Lotz04} showed that galaxies with elliptical light profiles have low asymmetries, those with spiral arms are more asymmetric and finally extremely irregular and merging galaxies are typically highly asymmetric. The last coefficient, smoothness (S), was introduced by \citet{Conselice03}. Smoothness is a good indicator of clumpiness of small scale structures of galaxies and it is related with star formation regions. \\
\indent The most widespread non-parametric coefficients used is the $Gini$ and the second-order moment of light distribution ($M_{20}$). The $Gini$ coefficient is based on the Lorenz curve, the rank ordered cumulative distribution function of a population\textquotesingle s wealth. It was proposed in 1912 by the Italian statistician and sociologist Corrado Gini, who used it to measure the inequality of the levels of income in a society. \citet{Abraham03}  extended the application of $Gini$ coefficient in the morphology of galaxies replacing the income of the society with the pixel values of a galaxy image. For the majority of local galaxies, the $Gini$ coefficient is correlated with C and increases with the fraction of light in a compact (central) component. $Gini$ is defined as: \\
\begin{equation} 
Gini=\frac{1}{|\bar{f_{i}}|k(k-1)}\sum_{i=1}^{k}{(2i-k-1)|f_{i}}| ,
\label{giniformula}
\end{equation}
where $\bar{f_{i}}$ is the mean flux value, $f_{i}$ is the flux of the i-pixel and $k$ is the number of pixels assigned to the galaxy.  The values of $Gini$ range from 0 to 1.  When a galaxy has a uniform flux distribution of pixels, the $Gini$ coefficient is close to zero. In the other extreme case, when the light of the galaxy concentrates in just a few pixels, $Gini$ is close to unity.\\
\indent In addition, $M_{20}$ traces the spatial distribution of any bright nuclei, bars, spiral arms, and off-centre star clusters \citep{Lotz04}. The definition of the $M_{20}$ is given by the following formula. 
\begin{equation}
M_{20}=log(\frac{\sum{M_{i}}}{M_{total}}) ,
\label{m20formula}
\end{equation} while  
\begin{equation}
\sum{M_{i}} < 0.2 flux_{total} ,
\end{equation}\\
The total second-order moment $M_{total}$ is the flux in each pixel $f_{i}$ multiplied by the squared distance to the centre of the galaxy, summed over all pixels assigned to the galaxy: 
\begin{equation}
M_{total}=\sum{M_{i}}=\sum{f_{i}[(x_{i}-x_{c})^{2}+(y_{i}-y_{c})^{2}]} ,
\end{equation}
where $x_{c}$, $y_{c}$ is the galaxy\textquotesingle s centre and $f_{i}$ is the flux of pixel $x_{i}$, $y_{i}$. The centre is computed by finding $x_{c}$, $y_{c}$ such that $M_{total}$ is minimized.  A disk galaxy with bright regions in the spiral arms or a more spatially extended object corresponds to $M_{20}$ values close to zero. Converesly, a galaxy with a bright bulge, single or double nucleus (a more compact object), takes more negative $M_{20}$  values.\\
\indent Since our galaxies exhibit characteristics of merging in their light distributions and many bright regions would spread around the centre of NIR light, we decided to extend our analysis also calculating the $M_{50}$ non parametric coefficient. The limit of 50\% of total flux in the definition of $M_{50}$ could reveal more bright regions and possibly classify in a more accurate way our sample (see Appendix \ref{appa} for details). \\
\indent We calculate $Gini$ as follows: We sort the pixel values from minimum to maximum. The pixels are divided into two, equal in number, separate groups. The first 50\% are the faint pixels and the remaining 50\%  are the brighter ones. The summation has $k$ terms (where $k$ is the number of the pixels inside the segmentation map). The $Gini$ coefficient gives negative values in the faint pixels and positive to the bright ones. With this approach, $Gini$ is calculated summing the difference between the brightest and the faintest pixel, the second brightest and second faintest pixel, etc. The final step is to divide with the term of $\left| f_{i}\right| k(k-1)$ in order to normalise the result and get the $Gini$ coefficient. \\
\indent The steps for the $M_{20}$ calculation are the following:
We arrange the pixel values in decreasing order of flux and we calculate the term $M_{i}$ for every pixel of the galaxy.
We define the centre of the galaxy $x_{c}$, $y_{c}$ as the point which represent the weighted centre of light.
We calculate the numerator of the term inside the logarithm $\sum{M_{i}}$ adding the moments of light of every pixel until the summed flux of the pixels reach the 20\% of the total flux of the segmentation map.
Finally, we calculate the $M_{20}$ coefficient. As the argument of the logarithm decreases (few pixels are needed to reach the 20\% of the total flux) the $M_{20}$ coefficient becomes more negative and the galaxy is characterised as compact. In the opposite case, as the numerator does not have great difference with the denominator, the number is bigger, the logarithm approaches values close to zero and the galaxy has a more extended morphology.
The important terms which are relevant on $M_{20}$ calculation are the number of pixels required to reach the 20\% of the total flux, the distance of these pixels from the centre and the relative difference between the flux of the brightest pixel (usually it lies in the centre of the galaxy) and the flux of the fainter outer pixels of the galaxy.

%=================================================================================
%================================   ANALYSIS  ======================================
%=================================================================================
\section{Analysis}

%Describe the method
\subsection{Constructing the segmentation map}
\begin{figure}
{\includegraphics[width=\hsize]{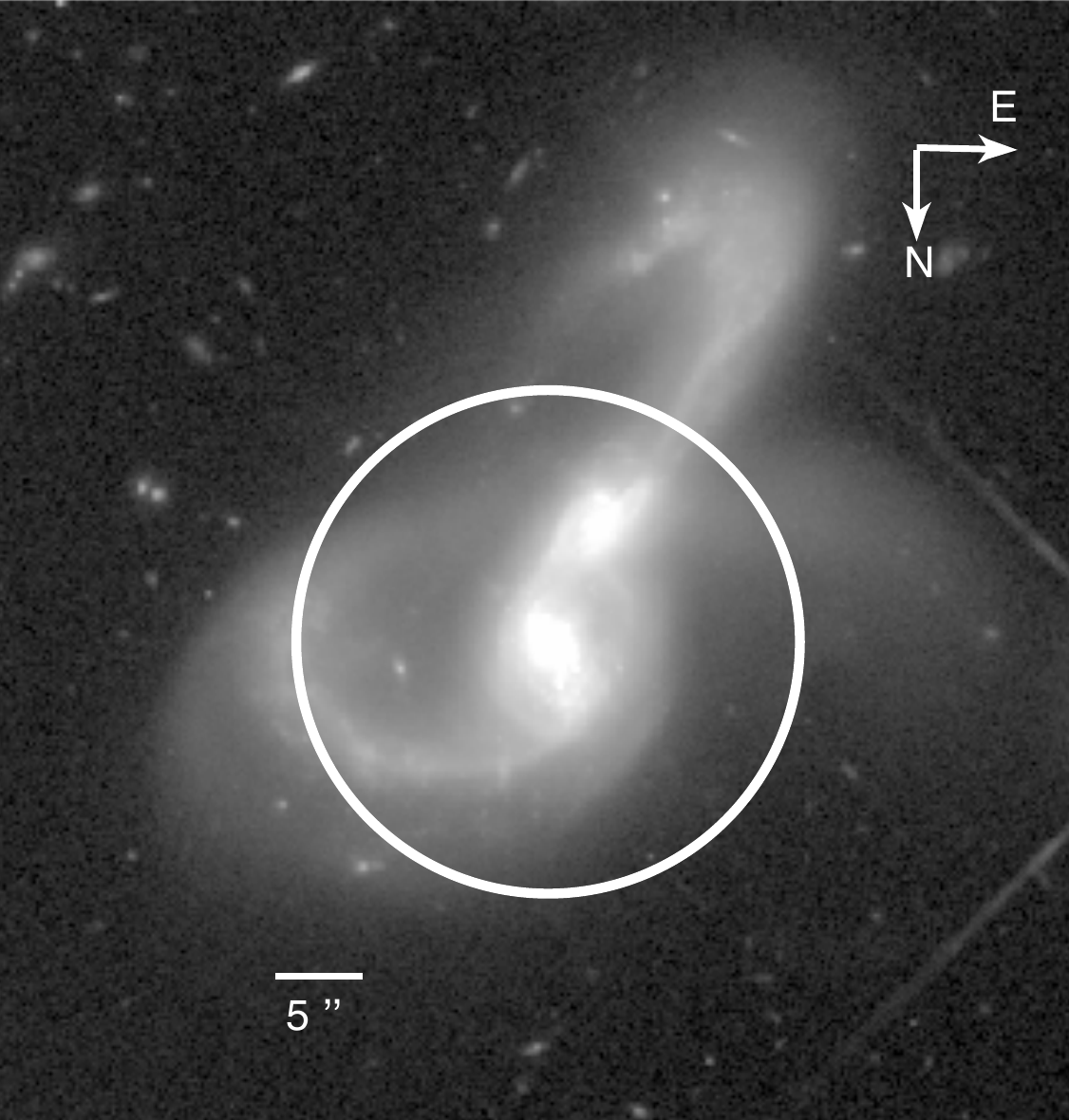}
           }        
            \resizebox{\hsize}{!}
           {\includegraphics[width=19cm,frame]{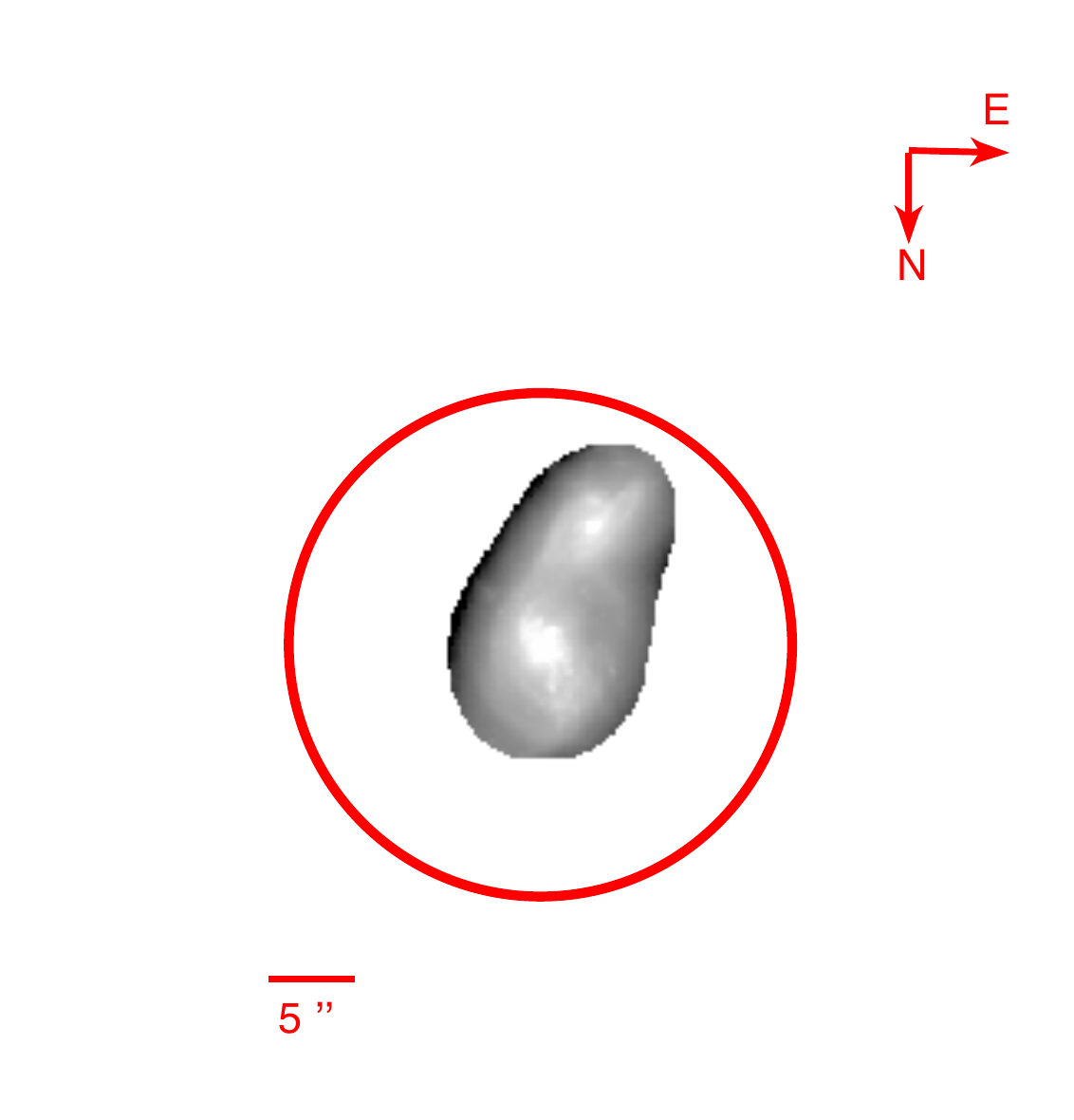}
           }
\caption{\small{The H-band image with the Petrosian radius represented as a white circle (top) and the the H-band segmentation map (bottom) with the Petrosian radius represented as a red circle of VV 705. The 2 images have the same FoV.}}
\label{segmentation_map_VV705}
\end{figure}

\citet{Lotz04} studied the $Gini$, $M_{20}$ values of a sample of local galaxies in both NUV and optical wavelengths. Their sample included galaxies with various morphological type, i.e. spirals, ellipticals, irregulars (Irr) and also (U)LIRGs. They identify the pixels that belong to each galaxy with a technique known as the construction of the segmentation map. The general idea is to set a flux threshold in every image of the galaxy. If the pixels of the image have a value above that limit we assume that they belong to the galaxy. The method requires a calculation of a characteristic radius of the galaxy. Most common are the definition of the Holmberg radius, the effective radius and the Petrosian radius \citep{Petrosian76}. Petrosian radius is based on a curve of growth and therefore is less affected by the $(1+z)^{4}$ surface brightness dimming of distant galaxies. For our analysis, we choose the Petrosian radius, which gives the opportunity to measure a characteristic radius of every galaxy independently of its distance. The equation that gives the Petrosian radius of a galaxy is the following :
\begin{equation}
\eta=\frac{\mu(r_{P})}{\bar{\mu}(r<r_{P})} ,
\end{equation}\\
where $\eta$ is typically set to 0.2.
The Petrosian radius ($r_{P}$) is the radius at which the surface brightness at $r_{P}$ is 20$\%$ of the mean surface brightness inside $r_{P}$. The surface brightness $\mu(r_{P})$ is measured for increasing circular apertures as the Petrosian radius determined by the curve of growth within circular apertures. We measure the flux inside an annulus and divide with the area of the annulus. The mean surface brightness $\bar{\mu}(r<r_{P})$ is the total flux inside an aperture divided by the area of the aperture.\\
\indent Firstly, the image must be sky subtracted and external sources like field stars or other galaxies must be removed. We define circular apertures around the brightest pixel in H-band and calculate the associated Petrosian radius. Furthermore, we convolve the cleaned galaxy image with a Gaussian of standard deviation $\sigma= (r_{P}/5)$, similarly to \citet{Lotz04}, in order to better trace low surface brightness pixels. The pixels assigned to the galaxy must satisfy two conditions. Firstly, their flux must be greater than the $\mu(r_{P})$ in the smoothed image. Secondly, we define a 3$\times$3 pixel area as the neighbourhood for every pixel, we measure the flux for every neighbour and if the flux difference between the central and every neighbouring pixel is less than 10 $\sigma$ then we add the pixel to the segmentation map. Finally, the map is applied to the cleaned but unsmoothed image and the pixels assigned to the galaxy are used to compute the $Gini$ and $M_{20}$  coefficients. In Fig. \ref{segmentation_map_VV705} we show the initial image and the segmentation map of NGC 34 in the H-band.\\
\indent For reasons of consistency in the morphological classification across all bands we decided to use in our analysis the same FoV at all wavelengths. Since the H-band images have the smallest FoV, they determine the area over which the nonparametric coefficients will be calculated and hence the morphology of the galaxy will be determined. As a first step the B- and I-band images where cropped to the corresponding FoV of the H-band. Furthermore, in order to be directly comparable with the H-band pixel scale and H-band morphology we deconvolved each image with the band-specific PSF and convolved with the H-band PSF. \\
\indent It should be noted that for 34 galaxies in our sample the Petrosian radius is inside the H-band FoV, and therefore within all filter images (since the H-band data has the smallest FoV in our dataset), and its projected linear scale at the distance of the galaxies has a median value of 9 kpc. For cases where the Petrosian radius extends outside the H-band FoV, we set these pixel values equal to the mean background of the image, which is close to zero, and then we calculate the corresponding Petrosian radius. We present the $Gini$, $M_{20}$ values of all 89 galaxies in Table \ref{gini_m20_table}. We should stress that the B- and I-band values of the 55 galaxies of Table \ref{gini_m20_table} for which the corresponding Petrosian radii are larger than the reduced cropped field may not accurately represent the actual value of the galaxy as a whole in this band. For this reason we mark them with a dagger symbol. Moreover, we provide for the reader in the Appendix \ref{appa} a complete Table with the $Gini$ and $M_{20}$ values of these galaxies in the B- and I-band calculated using exactly the same methodology but on the original uncropped FoV of each band.

%=================================================================================
%================================   RESULTS  =======================================
%=================================================================================
\section{Results}

%1st project
%=======================  Gini-M20 at different Petrosian radius ================================
\subsection{$Gini$ and $M_{20}$ at different Petrosian radius}
Given the limitations of the small FoV of the H-band, we decided to examine how the two non-parametric coefficients, $Gini$ and $M_{20}$ vary as a function of the Petrosian radii. \\
\indent We used the I-band as a reference since it has the best angular resolution and it reveals many details and features such as bars, tidal tails and bright regions. We construct the final image using four different Petrosian radius (0.67, 1, 1.5, 2) following the same method described earlier and we measure the $Gini$ and $M_{20}$ values in the four different segmentation maps.\\
\indent In Fig. \ref{variable_petrosian_I_abs_median}, we show the average $Gini$-$M_{20}$ of every morphological class of LIRG changes as the Petrosian radius increases. In order to check the direction of the $Gini$-$M_{20}$ loci of each LIRG as the Petrosian radius raises, we normalise all $Gini$-$M_{20}$ values according to the smallest value of $Gini$ and the largest value of $M_{20}$ in the sample. As a result, all LIRGs have the same point of origin (0,0) as the radius increases from 0.67 to 2 times the Petrosian value. Fig. \ref{variable_petrosian_I_abs_median} shows that there is a general trend for the majority of galaxies to raise their $Gini$ and get more negative $M_{20}$ values for larger radius. A possible explanation for that result could be the following:

\begin{figure}[h] \resizebox{\hsize}{!}{\includegraphics{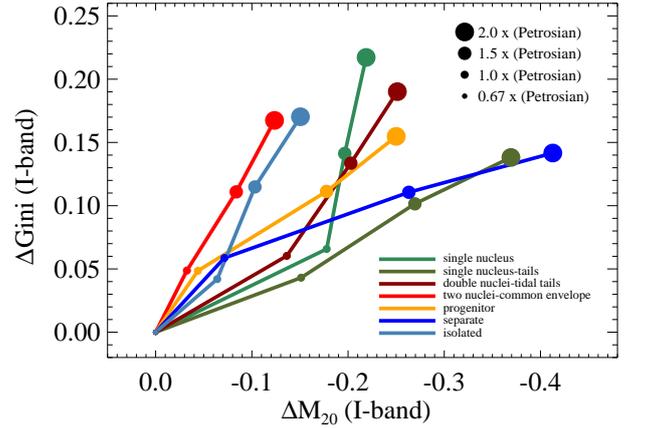}} 
\caption{\small{Plot presenting the change in $Gini$ and $M_{20}$ for various stages of interaction when the radius to calculate the segmentation map increases from 0.67 to 2 times the Petrosian value. $\Delta$$Gini$ (and $\Delta$$M_{20}$) is the difference between the $Gini$ (and $M_{20}$) at a given radius minus the value at the smallest radius. The increase of Petrosian radius is indicated by the size of the circles. The seven lines indicate the different morphological classification of LIRGs based on morphological classification of H11. In particular, sky blue, blue, orange, red, dark red, olive and green indicate isolated galaxies, separate galaxies (disks symmetric and no tidal tails), progenitor galaxies distinguishable with disks asymmetric or amorphous and/or tidal tails, two nuclei in common envelope, double nuclei plus tidal tail, single or obscured nucleus with long prominent tails and single or obscured nucleus with disturbed central morphology and short faint tails.}}
\label{variable_petrosian_I_abs_median}
\end{figure}

As the value of the radius increases, pixels with low surface brightness values enter the segmentation map. The influence in the calculation of $Gini$ could be important because as more fainter pixels enter the segmentation map, the weight of the light distribution becomes more skewed towards the central regions. The behavior of $M_{20}$ is more complicated. If we increase the radius, the segmentation map would be more extended, and one would expect $M_{20}$ values to be progressively less negative, closer to zero. However, if the light of the 20\% of the brightest pixels comes from a region close to the nucleus, or close to the regions around the two nuclei for double systems, the galaxy as a whole would have a more centrally concentrated light distribution which results in more negative $M_{20}$ values.\\

%2nd project
%=======================  Gini-M20 from optical to NIR  ================================
\subsection{Quantifying the Morphology of optical and NIR images}
In this section we calculate the $Gini$ and $M_{20}$ values of our sample in the optical and infrared bands, and study how their position in the \citet{Lotz04} diagram relates to their morphological classification according to \citet{Haan11}. \\
\indent \citet{Lotz04,Lotz08} divided the $Gini$-$M_{20}$ space in three regions for a sample of local galaxies as well as for a sample of the HST Survey of the Extended Groth Strip (EGS) at 0.2 $\le$ z $\le$ 0.4 . These regions identify a galaxy as merger, elliptical (E) or $S_{a}$ or disk-like and Irr.\\ 

\onecolumn

\setlength{\textwidth}{6.8in}
\LTcapwidth=\textwidth
%\scriptsize

\begin{longtable}{lcccccccc}
\hline 
\hline 
\caption[]{$Gini$, $M_{20}$ values of LIRGs in the B, I, H and IRAC 5.8$\mu$m band.} \\
\hline  % \\[-2.0ex]
 Optical ID & $ Gini$ (B) & $Gini$ (I) & $Gini$ (H) & $Gini$ (5.8$\mu$m) & $M_{20}$ (B) & $M_{20}$ (I) & $M_{20}$ (H) & $M_{20}$ (5.8 $\mu$m)  \\
 (1)&(2)&(3)&(4)&(5)&(6)&(7)&(8)&(9) \\
\hline 
\noalign{\smallskip}
\endfirsthead
\hline
\hline
\noalign{\smallskip}
\caption{continued.}\\
\hline 
\noalign{\smallskip}
Optical iD & $ Gini$ (B) & $Gini$ (I) & $Gini$ (H) & $Gini$ (5.8$\mu$m) & $M_{20}$ (B) & $M_{20}$ (I) & $M_{20}$ (H) & $M_{20}$ (5.8 $\mu$m) \\
(1)&(2)&(3)&(4)&(5)&(6)&(7)&(8)&(9) \\
\hline 
\noalign{\smallskip}
\endhead
\hline
\endfoot
\hline
\noalign{\smallskip}
\endlastfoot
\smallskip
NGC0034\textsuperscript{$\dagger$}  	         &  0.38  & 0.47 & 0.57 &    0.52  &  -1.60 & -2.23 & -2.01   &  -1.58  \\
ARP256N\textsuperscript{$\dagger$}   	         &  0.44  & 0.40 & 0.37 &    0.79  &  -0.75 & -1.01 & -0.97   &  -1.32  \\
ARP256S 	         &  0.43  & 0.41 & 0.51 &    0.56  &  -0.87 & -1.32 & -2.36   &  -1.73  \\
MCG+12-02-001\textsuperscript{$\dagger$}  	         &  0.45  & 0.42 & 0.49 &    0.55  &  -1.46 & -1.82 & -1.89   &  -1.62  \\
IC-1623\textsuperscript{$\dagger$}   	         &  0.58  & 0.51 & 0.38 &    0.56  &  -0.97 & -0.84 & -1.03   &  -1.30  \\
MCG-03-04-014	         &  0.37  & 0.37 & 0.43 &    0.50  &  -1.21 & -1.83 & -1.87   &  -1.67  \\
CGCG436-030\textsuperscript{$\dagger$}  	         &  0.33  & 0.36 & 0.56 &    0.53  &  -1.17 & -1.48 & -2.64   &  -1.78  \\
IRASF01364-1042          &  0.32  & 0.32 & 0.45 &    0.55  &  -1.16 & -1.51 & -2.07   &  -1.67  \\
IIIZw035	         &  0.56  & 0.50 & 0.57 &    0.51  &  -1.57 & -1.92 & -1.91   &  -1.81  \\
NGC0695\textsuperscript{$\dagger$}   	         &  0.39  & 0.36 & 0.39 &    0.46  &  -1.16 & -1.46 & -1.80   &  -1.55  \\
PGC9071 	         &  0.43  & 0.43 & 0.42 &    0.52  &  -1.19 & -1.92 & -2.03   &  -1.71  \\
PGC9074 \textsuperscript{$\dagger$}  	         &  0.46  & 0.48 & 0.49 &    0.56  &  -1.49 & -2.22 & -2.26   &  -1.87  \\
UGC02369S\textsuperscript{$\dagger$}  	         &  0.56  & 0.47 & 0.52 &    0.54  &  -1.58 & -1.59 & -1.68   &  -1.40  \\
IRASF03359+1523\textsuperscript{$\dagger$}            &  0.79  & 0.72 & 0.43 &    0.49  &  -1.44 & -0.83 & -0.62   &  -1.72  \\
ESO550-IG02\textsuperscript{$\dagger$}  	         &  0.44  & 0.43 & 0.41 &    0.70  &  -1.25 & -1.82 & -2.26   &  -0.77  \\
NGC1614\textsuperscript{$\dagger$}   	         &  0.52  & 0.56 & 0.49 &    0.59  &  -0.87 & -1.49 & -1.87   &  -1.87  \\
ESO203-IG001	         &  0.51  & 0.57 & 0.45 &    0.53  &  -0.69 & -1.00 & -1.85   &  -1.50  \\
VII-Zw-031\textsuperscript{$\dagger$}  	         &  0.37  & 0.35 & 0.41 &    0.48  &  -1.15 & -1.76 & -1.93   &  -1.56  \\
ESO255-IG007N\textsuperscript{$\dagger$}  	         &  0.67  & 0.49 & 0.41 &    0.50  &  -0.93 & -0.72 & -1.48   &  -1.82  \\
ESO255-IG007S\textsuperscript{$\dagger$}  	         &  0.53  & 0.52 & 0.49 &  $-$  &  -0.91 & -1.14 & -1.33   &   $-$	\\
AM0702-601N	         &  0.34  & 0.40 & 0.50 &    0.59  &  -1.46 & -2.16 & -1.41   &  -1.65  \\
AM0702-601S	         &  0.54  & 0.50 & 0.40 &    0.52  &  -1.62 & -1.56 & -1.32   &  -1.64  \\
2MASX-J07273754-0254540  &  0.50  & 0.40 & 0.38 &    0.52  &  -1.02 & -1.21 & -1.22   &  -1.89  \\
IRAS08355-4944           &  0.56  & 0.46 & 0.42 &    0.50  &  -1.50 & -1.36 & -1.09   &  -1.85  \\
NGC2623\textsuperscript{$\dagger$}   	         &  0.32  & 0.33 & 0.51 &    0.54  &  -1.18 & -1.65 & -2.47   &  -1.83  \\
ESO060-IG016\textsuperscript{$\dagger$}  	         &  0.41  & 0.59 & 0.60 &    0.79  &  -1.25 & -0.96 & -0.91   &  -1.95  \\
IRASF08572+3915\textsuperscript{$\dagger$}            &  0.44  & 0.45 & 0.52 &    0.52  &  -1.11 & -0.96 & -1.23   &  -1.73  \\
2MASX-J09133888-1019196\textsuperscript{$\dagger$}    &  0.49  & 0.45 & 0.52 &    0.53  &  -0.74 & -0.71 & -0.87   &  -1.74  \\
UGC04881	         &  0.40  & 0.43 & 0.47 &    0.64  &  -0.85 & -0.89 & -0.79   &  -1.00  \\
UGC05101\textsuperscript{$\dagger$}  	         &  0.32  & 0.41 & 0.57 &    0.52  &  -1.53 & -1.92 & -2.20   &  -1.71  \\
IRASF10173+0828          &  0.55  & 0.54 & 0.54 &    0.56  &  -1.76 & -1.95 & -2.11   &  -1.89  \\
NGC3256\textsuperscript{$\dagger$}   	         &  0.53  & 0.43 & 0.39 &    0.56  &  -1.42 & -1.54 & -1.69   &  -1.71  \\
IRASF10565+2448\textsuperscript{$\dagger$}            &  0.52  & 0.55 & 0.45 &    0.50  &  -0.79 & -2.07 & -1.94   &  -1.77  \\
ARP-148 	         &  0.58  & 0.51 & 0.53 &    0.52  &  -0.98 & -0.97 & -1.14   &  -1.77  \\
IRASF11231+1456\textsuperscript{$\dagger$}            &  0.43  & 0.43 & 0.50 &    0.63  &  -1.11 & -1.58 & -2.17   &  -2.17  \\
NGC3690W\textsuperscript{$\dagger$}  	         &  0.50  & 0.50 & 0.51 &    0.61  &  -0.74 & -1.02 & -0.83   &  -0.87  \\
NGC3690E\textsuperscript{$\dagger$}  	         &  0.35  & 0.28 & 0.52 &    0.73  &  -0.95 & -1.13 & -2.00   &  -0.89  \\
IRASF12112+0305\textsuperscript{$\dagger$}            &  0.42  & 0.46 & 0.53 &    0.57  &  -0.86 & -1.08 & -0.84   &  -1.22  \\
WKK0787\textsuperscript{$\dagger$}   	         &  0.47  & 0.50 & 0.60 &    0.53  &  -1.49 & -1.70 & -2.14   &  -1.75  \\
VV283		         &  0.44  & 0.54 & 0.50 &    0.53  &  -1.68 & -1.81 & -2.00   &  -1.73  \\
ESO507-G070\textsuperscript{$\dagger$}  	         &  0.41  & 0.49 & 0.56 &    0.53  &  -1.40 & -1.63 & -1.70   &  -1.74  \\
WKK2031\textsuperscript{$\dagger$}   	         &  0.32  & 0.52 & 0.36 &    0.52  &  -0.93 & -2.23 & -1.71   &  -1.80  \\
UGC08335W\textsuperscript{$\dagger$}  	         &  0.43  & 0.48 & 0.56 &    0.52  &  -1.38 & -1.65 & -1.89   &  -1.73  \\
UGC08335E	         &  0.55  & 0.44 & 0.57 &    0.49  &  -1.97 & -1.91 & -2.19   &  -1.56  \\
UGC08387\textsuperscript{$\dagger$}  	         &  0.31  & 0.32 & 0.45 &    0.51  &  -0.85 & -1.47 & -1.52   &  -1.73  \\
NGC5256\textsuperscript{$\dagger$}   	         &  0.39  & 0.41 & 0.46 &    0.74  &  -0.84 & -0.84 & -0.74   &  -1.39  \\
NGC5257\textsuperscript{$\dagger$}   	         &  0.35  & 0.25 & 0.37 &    0.46  &  -0.61 & -0.80 & -1.61   &  -1.00  \\
NGC5258\textsuperscript{$\dagger$}   	         &  0.54  & 0.53 & 0.55 &    $-$  &  -1.10 & -1.42 & -1.39   &   $-$	\\
UGC08696\textsuperscript{$\dagger$}  	         &  0.40  & 0.39 & 0.44 &    0.51  &  -1.18 & -1.61 & -1.50   &  -1.73  \\
NGC5331S\textsuperscript{$\dagger$}  	         &  0.39  & 0.41 & 0.53 &    0.55  &  -1.09 & -1.53 & -1.90   &  -1.09  \\
NGC5331 	         &  0.46  & 0.45 & 0.46 &    0.57  &  -1.27 & -1.45 & -1.96   &  -1.04  \\
IRASF14348-1447\textsuperscript{$\dagger$}            &  0.53  & 0.52 & 0.47 &    0.54  &  -0.98 & -1.24 & -0.95   &  -1.34  \\
IRASF14378-3651          &  0.37  & 0.43 & 0.58 &    0.53  &  -1.55 & -2.13 & -2.03   &  -1.83  \\
UGC09618S\textsuperscript{$\dagger$}  	         &  0.37  & 0.38 & 0.38 &    0.80  &  -1.40 & -1.80 & -1.97   &  -1.40  \\
VV705		         &  0.51  & 0.55 & 0.54 &    0.49  &  -1.31 & -1.38 & -1.03   &  -1.42  \\
ESO099-G004\textsuperscript{$\dagger$}  	         &  0.41  & 0.38 & 0.55 &    0.61  &  -0.58 & -0.90 & -0.82   &  -1.34  \\
IRASF15250+3608          &  0.40  & 0.35 & 0.44 &    0.54  &  -1.53 & -1.53 & -1.67   &  -1.94  \\
UGC09913\textsuperscript{$\dagger$}  	         &  0.24  & 0.23 & 0.46 &    0.52  &  -0.89 & -1.20 & -1.83   &  -1.61  \\
NGC6090\textsuperscript{$\dagger$}   	         &  0.64  & 0.55 & 0.59 &    0.52  &  -0.93 & -1.11 & -1.14   &  -1.48  \\
2MASXJ16191179-0754026\textsuperscript{$\dagger$}     &  0.37  & 0.41 & 0.39 &    0.52  &  -1.16 & -1.25 & -1.86   &  -1.66  \\
ESO069-IG006N\textsuperscript{$\dagger$}  	         &  0.41  & 0.49 & 0.40 &    0.50  &  -1.22 & -1.51 & -1.47   &  -1.55  \\
ESO069-IG006S	         &  0.41  & 0.51 & 0.51 &    0.58  &  -1.97 & -1.91 & -1.69   &  -1.78  \\
IRASF16399-0937\textsuperscript{$\dagger$}            &  0.45  & 0.36 & 0.50 &    0.48  &  -0.86 & -1.00 & -0.87   &  -1.46  \\
NGC6240\textsuperscript{$\dagger$}   	         &  0.29  & 0.40 & 0.57 &    0.53  &  -0.91 & -1.85 & -1.38   &  -1.84  \\
IRASF17132+5313          &  0.46  & 0.53 & 0.51 &    0.69  &  -0.94 & -0.81 & -0.56   &  -1.52  \\
IRASF17138-1017\textsuperscript{$\dagger$}            &  0.30  & 0.32 & 0.37 &    0.56  &  -0.86 & -1.22 & -1.22   &  -1.39  \\
IRASF17207-0014          &  0.32  & 0.32 & 0.46 &    0.53  &  -0.99 & -1.32 & -1.70   &  -1.55  \\
IRAS18090+0130           &  0.39  & 0.42 & 0.47 &    0.55  &  -1.04 & -1.78 & -2.08   &  -1.95  \\
IC4689S 	         &  0.42  & 0.39 & 0.40 &    0.53  &  -1.12 & -1.51 & -2.04   &  -1.69  \\
IRASF18293-3413\textsuperscript{$\dagger$}            &  0.52  & 0.46 & 0.47 &    0.50  &  -0.65 & -1.25 & -1.77   &  -1.67  \\
NGC6670B\textsuperscript{$\dagger$}  	         &  0.69  & 0.62 & 0.55 &    0.78  &  -0.93 & -1.00 & -1.41   &  -0.86  \\
NGC6670A\textsuperscript{$\dagger$}  	         &  0.48  & 0.50 & 0.61 &    0.80  &  -1.24 & -1.92 & -2.02   &  -0.87  \\
NGC6786S\textsuperscript{$\dagger$}  	         &  0.48  & 0.43 & 0.44 &    0.53  &  -1.76 & -2.05 & -2.15   &  -1.78  \\
ESO593-IG008	\textsuperscript{$\dagger$}           &  0.44  & 0.40 & 0.45 &    0.56  &  -0.80 & -1.15 & -1.43   &  -1.60  \\
IRASF19297-0406          &  0.43  & 0.57 & 0.46 &    0.52  &  -1.22 & -1.71 & -1.25   &  -1.75  \\
IRAS19542+1110           &  0.48  & 0.50 & 0.39 &    0.53  &  -1.78 & -1.77 & -1.79   &  -1.72  \\
IRAS20351+2521\textsuperscript{$\dagger$}             &  0.35  & 0.36 & 0.53 &    0.50  &  -0.89 & -1.72 & -1.90   &  -1.85  \\
IIZW096S\textsuperscript{$\dagger$}  	         &  0.44  & 0.60 & 0.37 &    0.67  &  -1.15 & -0.80 & -0.70   &  -1.20  \\
ESO286-IG019	         &  0.52  & 0.46 & 0.56 &    0.53  &  -1.24 & -1.91 & -2.00   &  -1.85  \\
IRAS21101+5810           &  0.40  & 0.43 & 0.45 &    0.52  &  -0.79 & -0.67 & -0.69   &  -1.65  \\
ESO239-IG002	\textsuperscript{$\dagger$}           &  0.43  & 0.47 & 0.59 &    0.54  &  -2.07 & -2.38 & -2.60   &  -1.81  \\
IRASF22491-1808          &  0.42  & 0.65 & 0.45 &    0.56  &  -1.00 & -1.66 & -1.00   &  -1.58  \\
ESO148-IG002	         &  0.38  & 0.42 & 0.53 &    0.62  &  -0.76 & -0.86 & -0.97   &  -1.36  \\
IC5298  	         &  0.43  & 0.47 & 0.49 &    0.51  &  -2.14 & -2.14 & -1.77   &  -1.79  \\
ESO077-IG014\textsuperscript{$\dagger$}  	         &  0.57  & 0.62 & 0.35 &    0.50  &  -0.87 & -0.95 & -1.23   &  -1.71  \\
NGC7674\textsuperscript{$\dagger$}   	         &  0.40  & 0.39 & 0.43 &    0.52  &  -1.71 & -2.45 & -2.67   &  -1.88  \\
IRASF23365+3604          &  0.32  & 0.39 & 0.48 &    0.56  &  -1.21 & -1.56 & -2.41   &  -1.77  \\
IRAS23436+5257           &  0.41  & 0.45 & 0.48 &    0.61  &  -1.13 & -1.01 & -0.82   &  -1.33  \\
UGC12812	         &  0.61  & 0.55 & 0.50 &    0.48  &  -1.25 & -1.65 & -1.66   &  -1.58  \\
\label{gini_m20_table}
\end{longtable}
\tablefoot{Columns: (1) Optical cross-identification, where available from NED (see \citet{Armus09} for details). (2),(3),(4) $Gini$ values calculated in the corresponding band using one Petrosian radius. (5) 5.8$\mu$m $Gini$ values calculated using two times the Petrosian radius. (6), (7), (8) $M_{20}$ values calculated in the corresponding band using one Petrosian radius. (9) 5.8$\mu$m $M_{20}$ values calculated using two times the Petrosian radius. Due to the coarser angular resolution of IRAC, there are two LIRG systems for which we were not able to obtain non-parametric coefficients for each individual galaxy resolved by $HST$. These are NGC5258 and ESO255-IG007S. The $Gini$ and $M_{20}$ values for these were thus measured for the entire system. As we discussed in Section 4.1, for 55 galaxies the cropped B- and I-band HST maps were smaller than the corresponding Petrosian radii. We mark those galaxies with a dagger, since the $Gini$ and $M_{20}$ values in the B- and I-band represent only the morphology of the fraction of the galaxy which fits within the footprint of the H-band. The $Gini$ and $M_{20}$ values of each galaxy using the whole B- and I-band field are presented in the Appendix \ref{appa}.}
\twocolumn

\begin{figure}%[!ht]%for 2-column-wide plot use figure*
           {\includegraphics[width=\hsize]{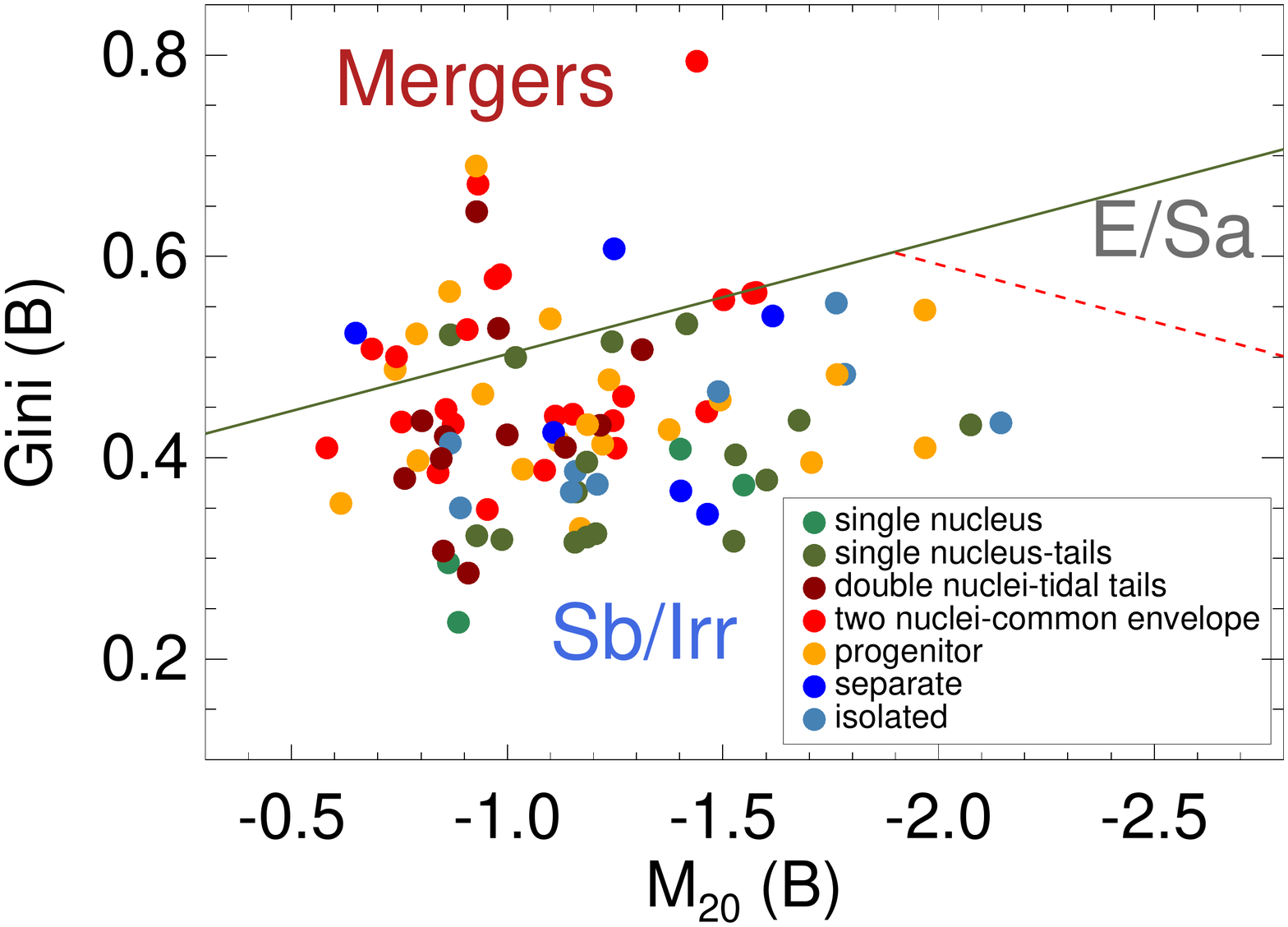}
            }        
            \resizebox{\hsize}{!}
           {\includegraphics[width=\hsize]{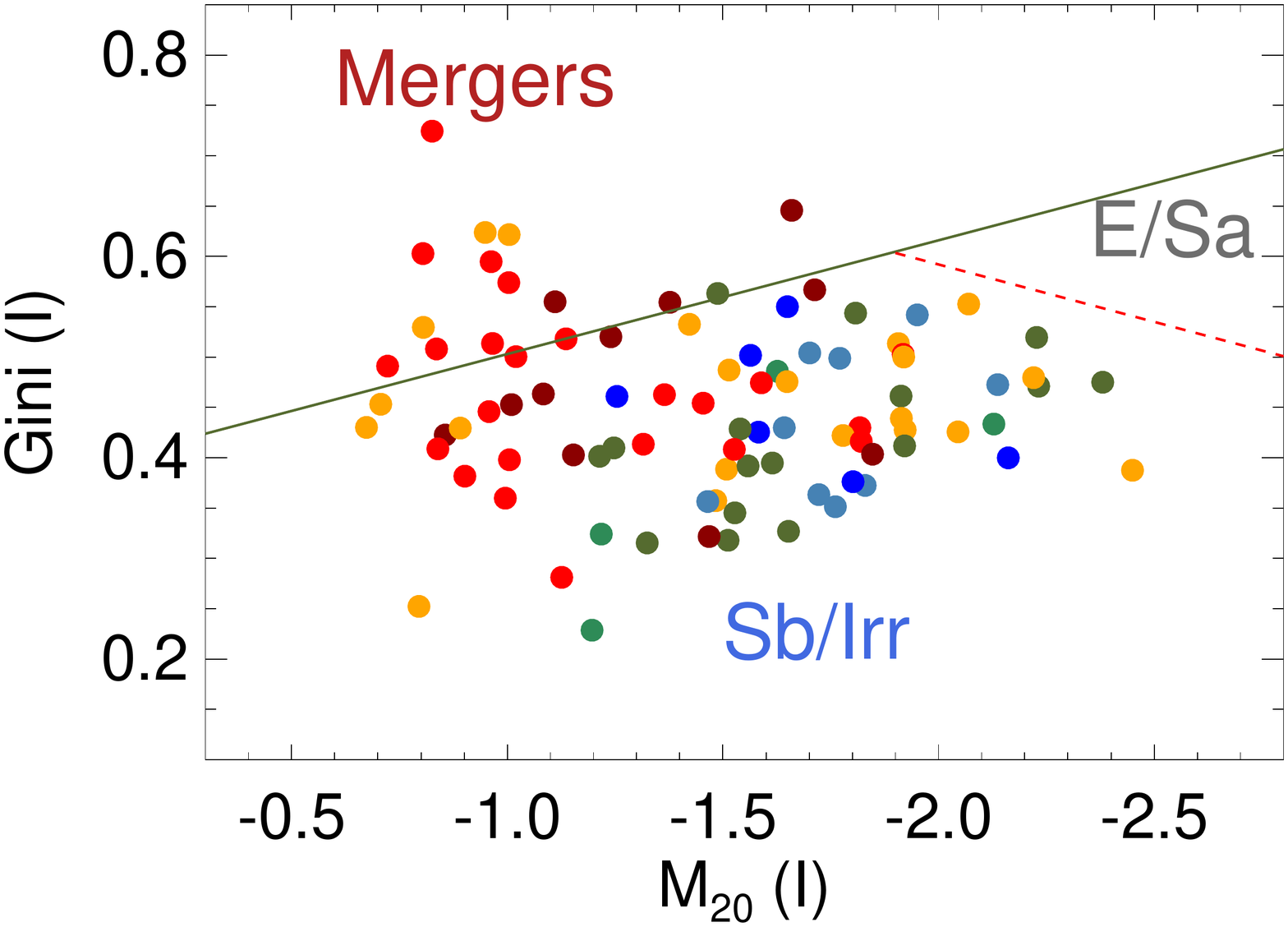}}
	           {\includegraphics[width=\hsize]{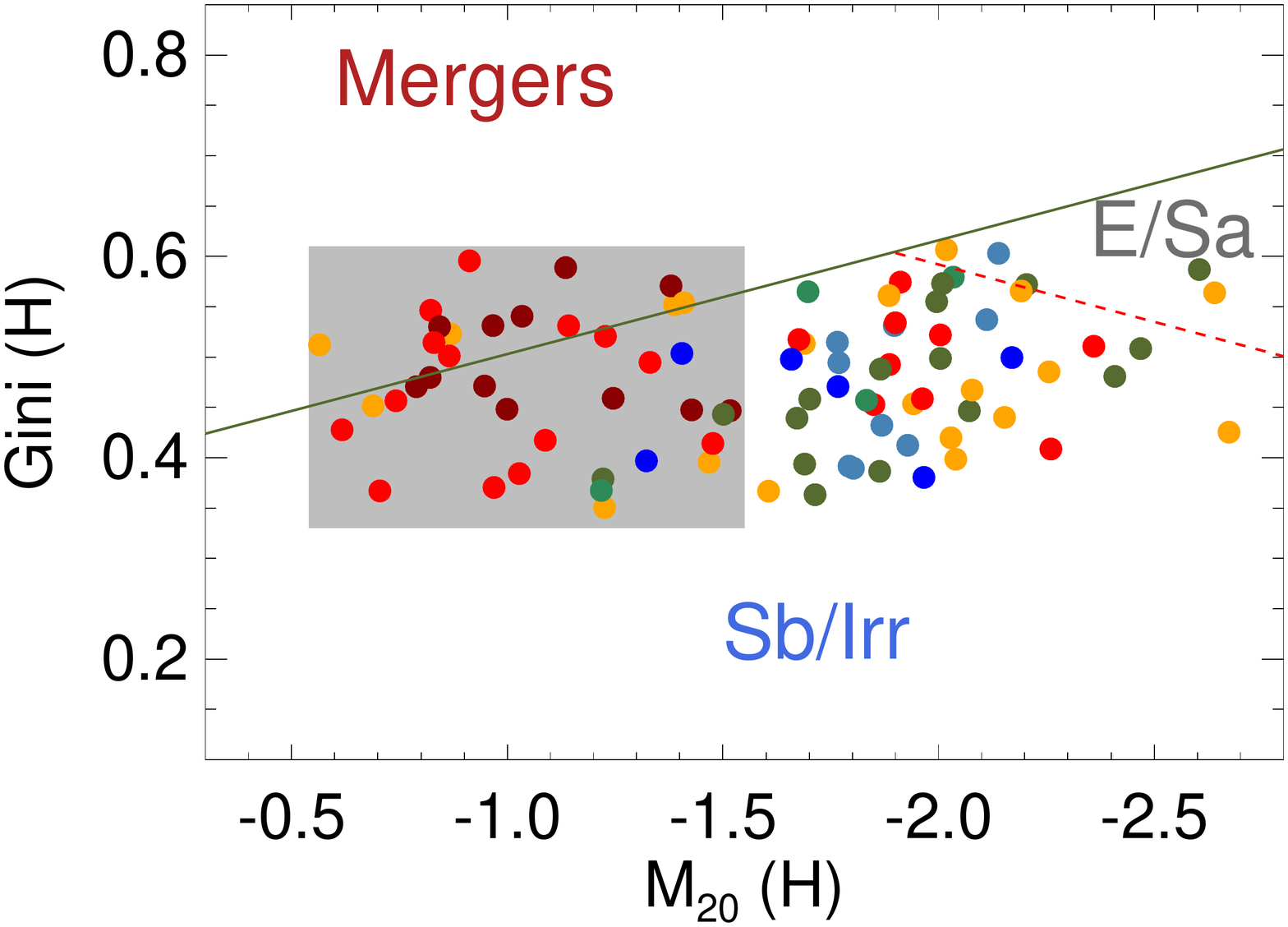}}
\caption{\small{$Gini$-$M_{20}$ space in B- (top), I- (middle) and H-band (bottom).
Colored filled circles indicate the different morphological classification of LIRGs following H11, where isolated galaxies, separate galaxies (disks symmetric and no tidal tails), progenitor galaxies distinguishable with disks asymmetric or amorphous and/or tidal tails, two nuclei in common envelope, double nuclei plus tidal tail, single or obscured nucleus with long prominent tails and single or obscured nucleus with disturbed central morphology and short faint tails respectively. Following \citet{Lotz04}, the upper green solid line separates merger candidates from normal Hubble types while the lower red dotted line divides normal early-types ($E/S_{a}$) from late-types ($S_{b}/Irr$). The grey rectangle in the H-band is the region where on-going mergers live regardless of the band. We argue that in H-band, this region can be used to better identify ongoing mergers.}}
\label{Gini_M20_H_center}
\end{figure}

\indent We expect that most of our galaxies will lie in the upper region of the $Gini$-$M_{20}$ parametric space because of their merger characteristics (such as tidal tails, double nuclei and morphologically disturbed structures).\\ 
\indent We also examine how the $Gini$-$M_{20}$ space is related with $M_{\star}$, $L_{IR}$ and SFR. We divide our sample in three bin according to their stellar mass ($M_{\star}$) having the same number of galaxies in every bin: (high $M_{\star}$ LIRGs : $M_{\star}>1.52x10^{11} M_{\sun}$, moderate $M_{\star}$ LIRGs : ($9.24x10^{10} M_{\sun}<M_{\star}<1.52x10^{11} M_{\sun}$) and small $M_{\star}$ LIRGs : $M_{\star}<9.24x10^{10} M_{\sun}$). In addition, we separate the sample as a function of $L_{IR}$ into sub-LIRGs, LIRGs and ULIRGs. Finally we separate the galaxies based to the SFR into low ( < 50 $M_{\sun}  yr^{-1}$ ), moderate ( 50 < $M_{\sun}  yr^{-1}$ < 100 ) and high ( > 100 $M_{\sun}  yr^{-1}$ ) SFR.  \\
\indent In Fig. \ref{Gini_M20_H_center} we present the $Gini$-$M_{20}$ space as a function of wavelength according to their H11 classification.

\newpage

In Fig. \ref{Gini_M20_H_center} we present the location of our LIRG sample in the $Gini$-$M_{20}$ space as a function of wavelength along with the H11 classification. We see that in the B-band, where the light comes from relatively unobscured young and intermediate age stars and star formation regions, 17 galaxies, out of 89 of the sample, are within the merger locus and none of them is in the region of $E/S_{a}$. In the I-band, where we expect more evolved stars to contribute to the light, 14 galaxies are below the merger line but still none of them is in the region of $E/S_{a}$. In the NIR, where we can probe structures deeper into the nuclei and the light comes from low mass main sequence K-stars not affected much by dust extinction, a few galaxies enter the $E/S_{a}$ region. We estimate the median values of the two parameters for every band. We identify a trend in that the median $Gini$ values increase and the median $M_{20}$ appear to decrease as we move from optical to NIR. Those results are in agreement with the study of \citet{Petty14}, who examined a smaller sample, also including UV observations. In addition, the difference between the maximum and the minimum of $Gini$ values decreases while for $M_{20}$ increases. We show these values in Table \ref{median_gini_m20}.\\

\begin{table}[h]
\caption{Median $Gini$ and $M_{20}$ values from optical to NIR wavelengths.}
\label{median_gini_m20}
\centering
\begin{tabular}{c c c c c}
\hline\hline
\smallskip
band & $Gini_{median}$ & $M_{20  median}$ & $Gini_{range}$ & $M_{20  range}$ \\
(1)&(2)&(3)&(4)&(5) \\
\hline
\smallskip
B & 0.43 & -1.15 & 0.56 & 1.56  \\
I & 0.45 & -1.51 & 0.50 & 1.78 \\
H & 0.49 & -1.70 & 0.26 & 2.11 \\
\hline
\end{tabular}
\tablefoot{Columns: (1) The reference band. (2) The median $Gini$ value. (3) The median $M_{20}$ value. (4) The $Gini$ range. (5) The $M_{20}$ range. }    %notes of the table
\end{table}

In general, as we characterise the morphology at longer wavelengths from B- to H- band, we see that isolated or pre-merger galaxies tend to reach more negative $M_{20}$ values while ongoing mergers tend to lie in the left region. We find that a significant fraction (36\%) of the sample do not change their location in the diagram as a function of wavelength and remain in the same region. More than 3/4 (78\%) of these LIRGs are double or triple systems classified as ongoing mergers and have $-0.56 \ge M_{20} \ge -1.55$ regardless of the band we are using to calculate the parameters. The low nearly, constant $M_{20}$ values are a consequence of the extent of the systems, which clearly show well separated members, regardless of how extended they are individually. This property can be used as a very useful tool in order to identify galaxies in this particular merger phase, specially in the NIR, where the contamination from systems in other interacting stages is minimal. That is, galaxies falling in this wedge of the parameter space (see the gray rectangle at the bottom of Fig. \ref{Gini_M20_H_center}) are most likely systems suffering an ongoing merger event. The latter result is consistent with \citet{Petty14} who found that quantitative $Gini$, $M_{20}$ measurements do not effectively reflect the wavelength dependence of merging systems. \\
\indent We also separate our sample according to $M_{\star}$ and check their positions inside the $Gini$-$M_{20}$ plane.

%4th project
%=======================  Gini-M20 at MIR  ================================
\subsection{The $Gini$ and $M_{20}$ classification at 5.8 $\mu$m }
In the following we explore the morphology of our galaxies as traced by the IRAC 5.8 $\mu$m filter. This choice was motivated by the fact that the filter samples both the 6.2$\mu$m Polycyclic Aromatic Feature, which traces star formation, as well as an underlying continuum mostly due to heated dust grains \citep{Smith07}.\\
\indent We calculate the non-parametric coefficients at 5.8 $\mu$m following the same method for creating a segmentation map as in the optical and NIR bands. In this case we use the MIR image to find the galaxy centre and the Petrosian radius. Finally, we construct the segmentation map of every galaxy using two times the Petrosian radius instead of one Petrosian radius we used in the previous paragraphs.

\begin{figure}[h] \resizebox{\hsize}{!}{\includegraphics{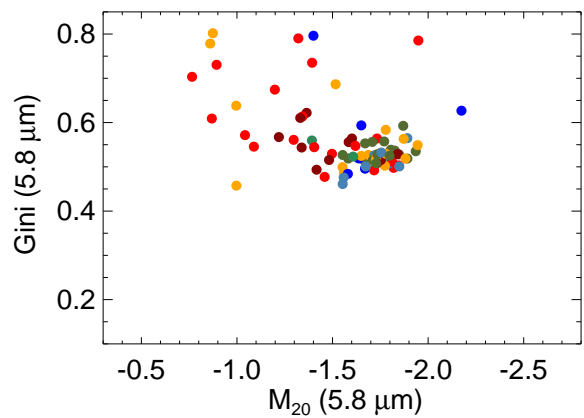}} 
\caption{\small{$G$-$M_{20}$ non-parametric space of IRAC 5.8$\mu$m. The points are marked following the colour scheme of Fig. \ref{Gini_M20_H_center}, to indicate the morphology classes according to H11.}}
\label{Gini_IRAC_M20_IRAC_bulge}
\end{figure}
Our results, presented in Fig. \ref{Gini_IRAC_M20_IRAC_bulge}, suggest that most of the galaxies of the sample is grouped in a clump on the $Gini$-$M_{20}$ plane with $Gini$ values $\sim$ 0.5 and $-1.5\geq M_{20} \geq -2.0$. The rest are scattered mostly towards the upper left of the plot at higher $Gini$ values. The reason for the clump is mainly due to the $\sim$20 fold decrease in angular resolution of the Spitzer images compared to the HST, which leads to a larger fraction of galaxies having the bulk of their emission originating from an unresolved central source. The remaining are systems which are either in early stage of interaction or harbouring resolved double nuclei, causing their corresponding $M_{20}$ values to be less negative.
%5th project
%=======================  Luminosity bins  ================================
\subsection{Luminosity bins of optical and NIR images}
It is important to examine how the distribution of our sample in the $Gini$-$M_{20}$ space varies according to $L_{IR}$ of galaxies. Since all ULIRGs ($L_{IR}$ > $10^{12}L_{\sun}$) in the local Universe are mergers, and therefore display disturbed morphologies (tidal tails, double nuclei etc.) \citep{SanMir96,Farrah01,Veilleux02,Ishida04}, we expected that all would lie above the merger line. 

\begin{figure}[!ht]%for 2-column-wide plot use figure*
           {\includegraphics[width=\hsize]{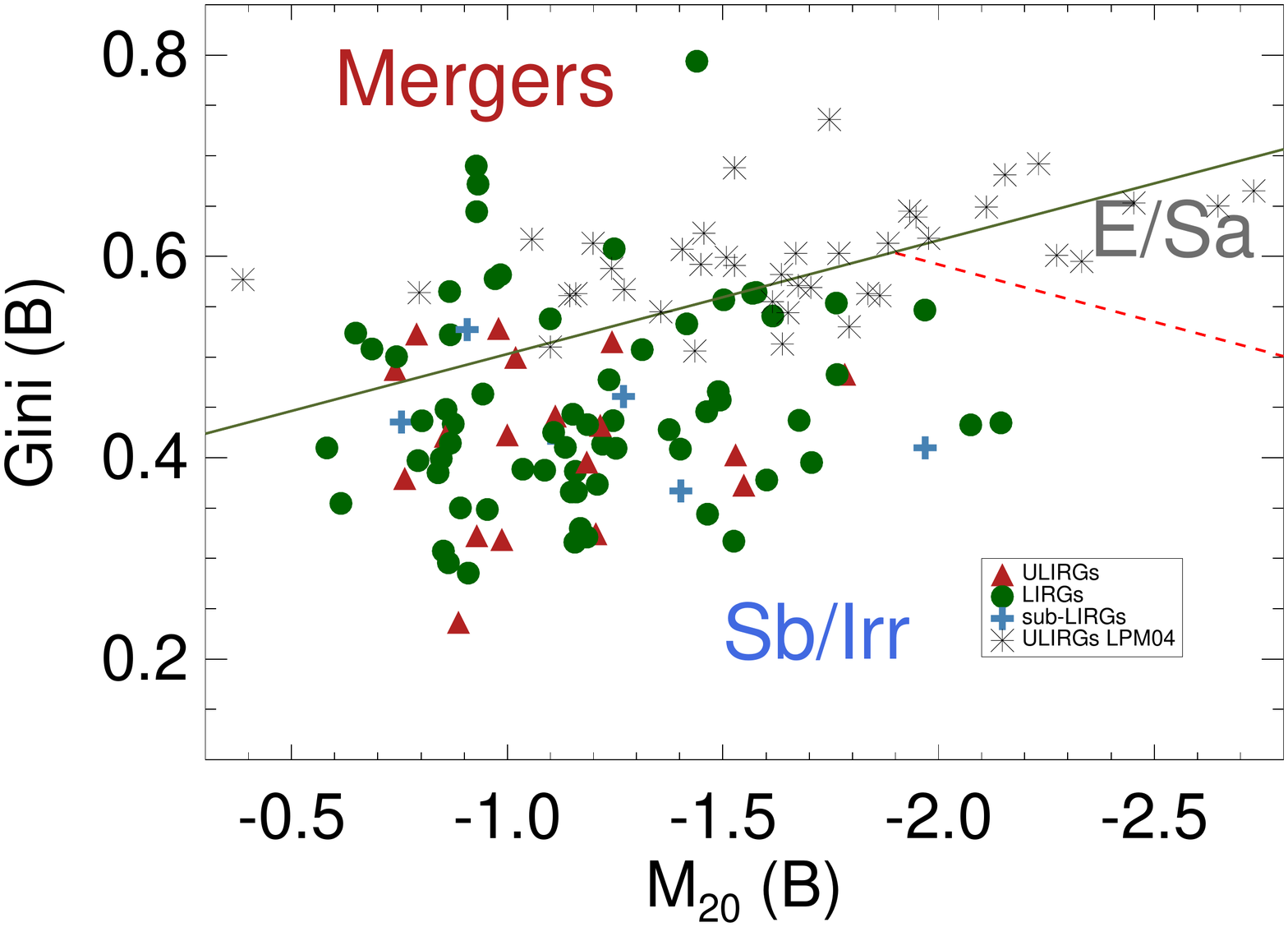}
           }        
            \resizebox{\hsize}{!}
           {\includegraphics[width=\hsize]{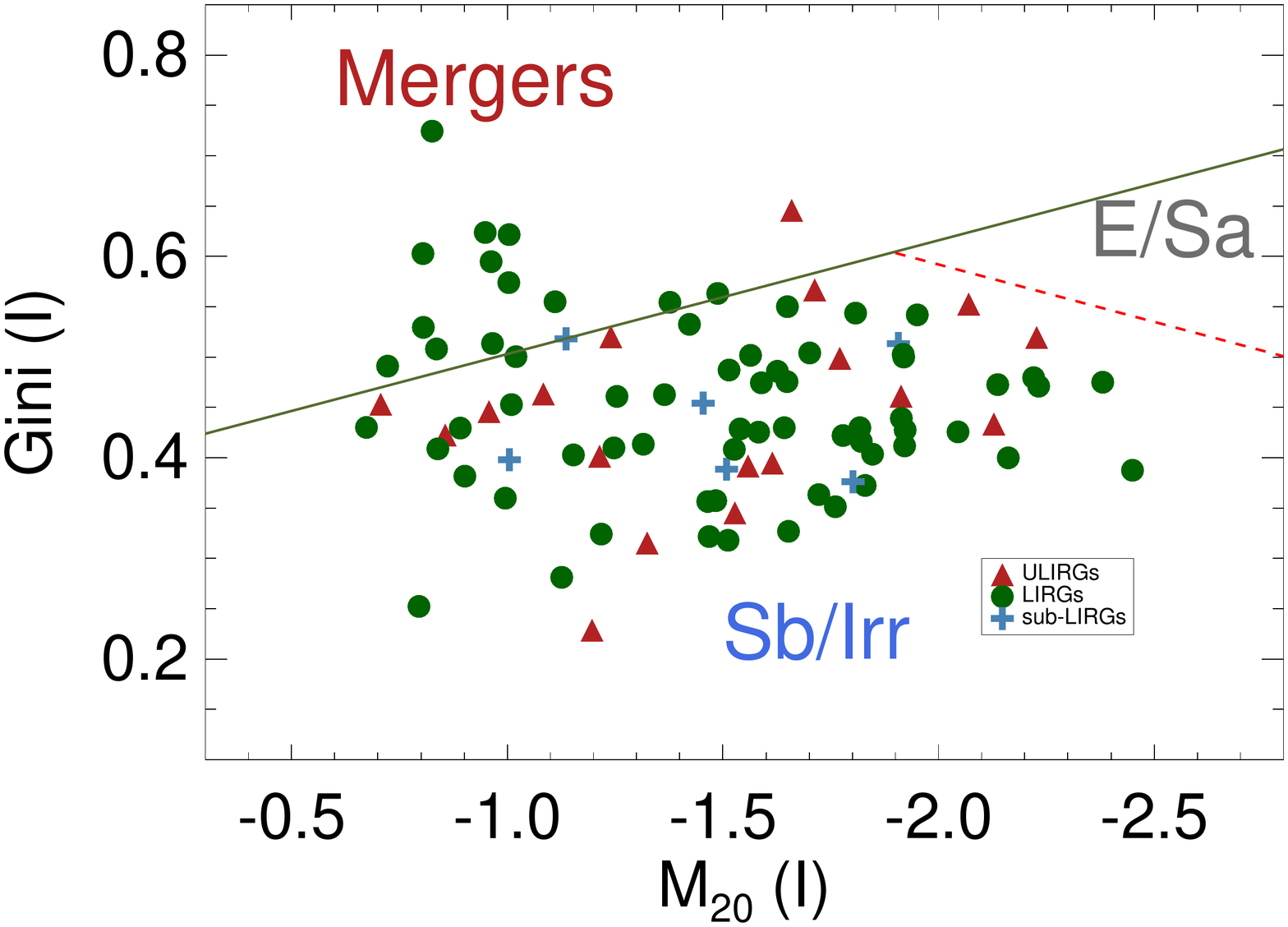}
           }
	   {\includegraphics[width=\hsize]{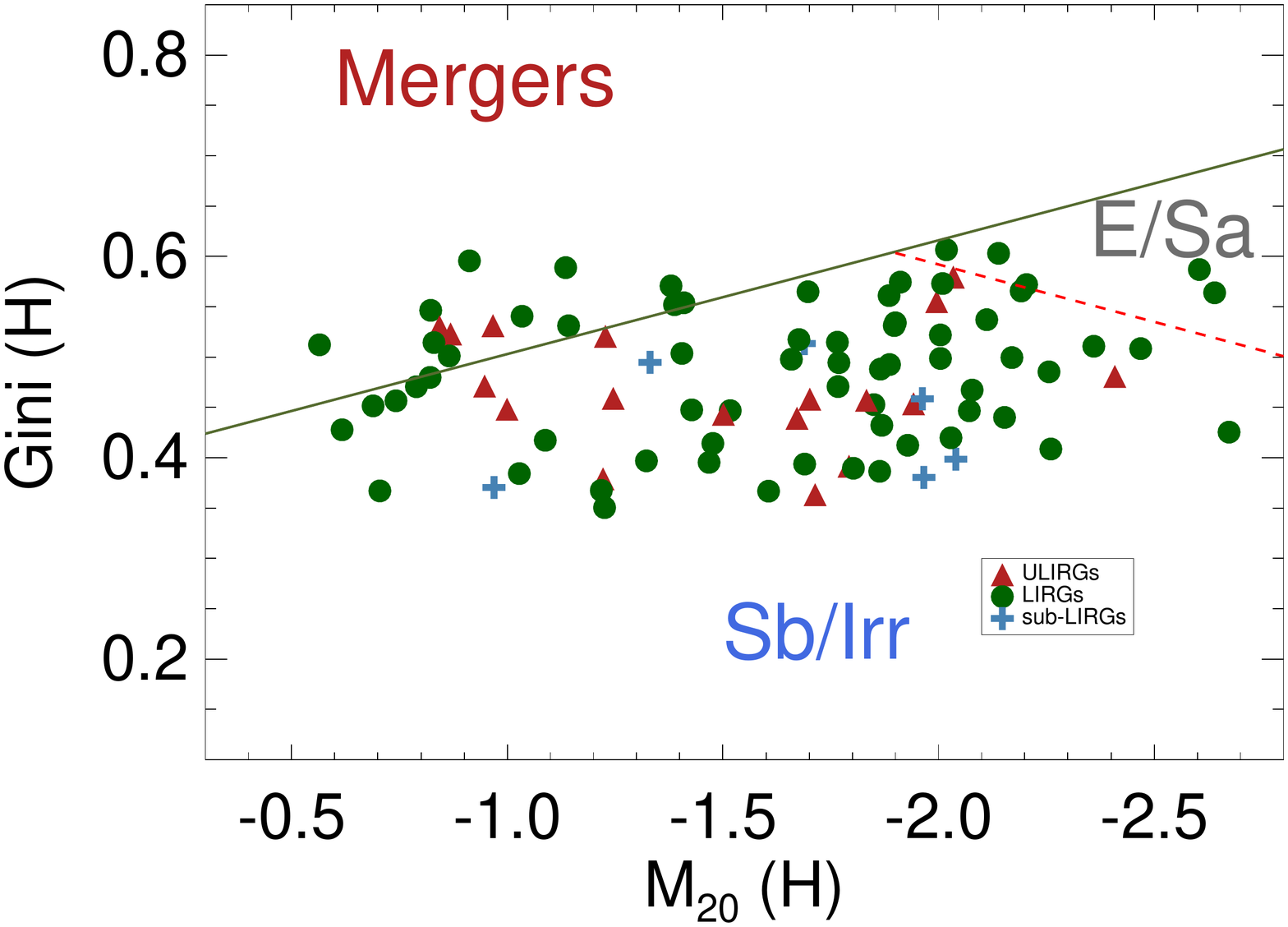}
	   }
\caption{\small{Same as in Fig. \ref{Gini_M20_H_center}, but this time the galaxies are grouped by their luminosity. The steel blue crosses indicate sub-LIRGs, the green filled circles LIRGs and the red triangles ULIRGs. In the B-band, we also show with black asterisks the ULIRG sample that \citet{Lotz04} used to define the merger region in the $Gini$-$M_{20}$ plane. Even though all ULIRGs in our sample are on-going mergers, they do not appear to populate the corresponding part of the $Gini$-$M_{20}$ plane.}}
\label{Gini_M20_H_center_lum_bins}
\end{figure}

\indent However, Fig. \ref{Gini_M20_H_center_lum_bins} shows that the ULIRGs in our sample have small $Gini$ values and most of them are under the merger line in all bands, which is not consistent with their visual appearance. This rather unexpected result is attributed to the fact that, as we mentioned in Section 4.1, we used the rather small FoV of the H-band images, as a reference also for the B and I-band observations. By estimating the $Gini$ and $M_{20}$ parameters within this region, which corresponds to a projected area of 13.4 $x$ 13.4 kpc at the median distance of 145 Mpc for our sample, several of the merging characteristics of the more nearby systems (in particular ULIRGs), such as long tidal tails and bridges, are being suppressed and do not contribute much to determining the locus of the galaxies in the $Gini$-$M_{20}$ plane. This is not the case for more distant systems or cases where double nuclei are clearly visible within the image. This was verified this by reevaluating the parameters for the full FoV of the B- and I-band images which reveals that these galaxies move towards the top left of the plane (see also \citet{Petty14}. Furthermore, Larson et al. (2016 in prep. ) also calculated non-parametric coefficients for part of our sample using only the I-band and a slightly different methodology in creating the segmentation map, and confirm this trend.

\newpage

%=======================  sSFR and M20  ================================
\subsection{Morphology and specific SFR }
\begin{figure}[h] \resizebox{\hsize}{!}{\includegraphics{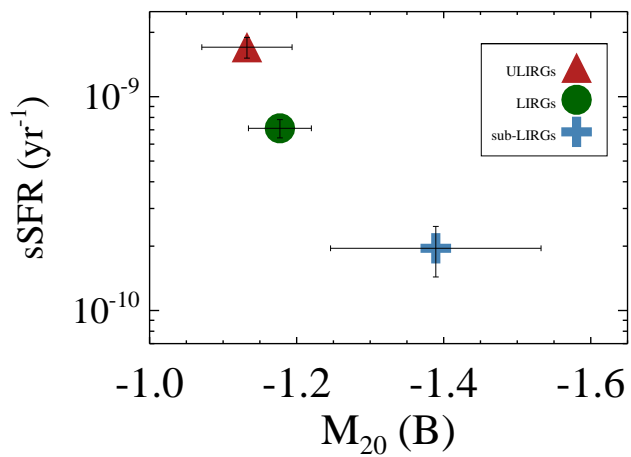}} 
\caption{\small{sSFR-$M_{20}$ in B-band. The steel blue cross corresponds to the mean value of sub-LIRGs sample, the green filled circle indicates the mean value of LIRGs and the red triangle represents the mean value of ULIRGs. The error bars are the standard deviation of the mean values. The larger standard deviation of the sub-LIRG point is due to the fact that there is a small number of galaxies (only $8\%$ of the whole sample) in this category.}}
\label{sSFR_hfov_vs_M20_B_FIELD_bulge_lum_bins_mean_err__B_field}
\end{figure}

\begin{figure}[h] \resizebox{\hsize}{!}{\includegraphics{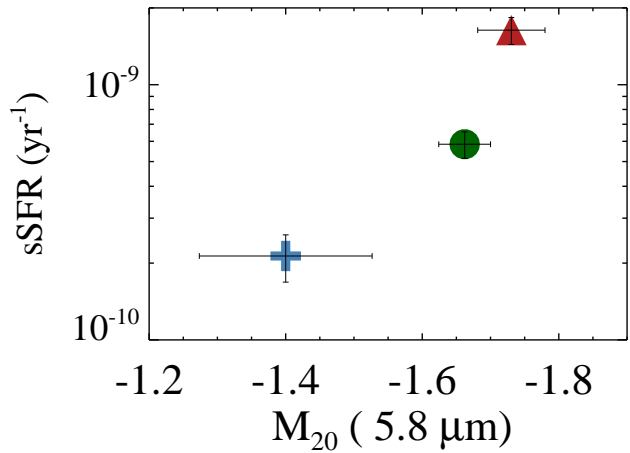}} 
\caption{\small{sSFR-$M_{20}$ in IRAC 5.8$\mu$m. The steel blue crosses correspond to sub-LIRGs, the green filled circles indicate LIRGs and finally the red triangles represent the ULIRGs. The error bars are the standard deviation of the mean values.}}
\label{sSFR_vs_M20_IRAC_bulge_lum_bins_med_err}
\end{figure}

We also examined whether there is a relation between the non-parametric coefficients and the sSFR=(SFR/$M_{\star}$). The stellar masses are calculated from IRAC 3.6 $\mu$m and 2MASS K-band photometry \citep{Lacey08,Howell10}. For some LIRGs without reliable K-band photometry, the masses are estimated from 3.6 $\mu$m data and scaled by the median ratio of (K-band)mass/(3.6 $\mu$m)mass from galaxies with measurements in both wavelengths. Our LIRGs have a stellar mass range of $2.54x10^{10} M_{\sun}$ < $M_{\star}$ < $8.15x10^{11} M_{\sun}$. The calculations of SFR were done following the equation of \citet{Kennicutt98} for starburst galaxies assuming that all $L_{IR}$ comes from reprocessing of star light, and the $L_{IR}$ values provided in \citet{DiazSantos13}. \\
\indent The main sequence (MS) of star-forming galaxies indicated by the SFR-$M_{\star}$ correlation can be interpreted by the fact that most galaxies spend most of their time producing stars at a normal pace, at least up to z$\sim$2 \citep{Elbaz07,Daddi07,Elbaz11}. Observations over a wide range of redshifts suggest that the slope of the SFR-$M_{\star}$ relation is almost unity (e.g. \citep{Elbaz07,Salmi12}, which implies that their sSFR does not depend strongly on stellar mass. \\
\indent In Fig. \ref{sSFR_hfov_vs_M20_B_FIELD_bulge_lum_bins_mean_err__B_field} we show the sSFR as a function of $M_{20}$ as measured in full ACS B-band maps (see Appendix \ref{appa}), where galaxies have been grouped in luminosity bins. We see that as $L_{IR}$ increases, the $M_{20}$ values as well as the sSFR become larger. That is, the B-band emission from ULIRGs, which mostly traces the unobscured stellar populations, appears more extended than in less starbursting galaxies. However, when we investigate the relation between sSFR and $M_{20}$ as measured using the IRAC 5.8 $\mu$m emission, we find the opposite trend (see Fig. \ref{sSFR_vs_M20_IRAC_bulge_lum_bins_med_err}). Galaxies with higher IR luminosities and sSFR become increasingly compact (show more negative $M_{20}$ values), in agreement with \citet{Elbaz11} and \citet{Diaz-Santos10}.

The most likely explanation for the most luminous (U)LIRGs being more extended in the B-band while appearing more compact at 5.8 $\mu$m is the spatial decoupling between the UV/optical and the MIR emission (see also \citet{Charmandaris04,Howell10}). In other words, the nuclei of the most compact (U)LIRGs become optically thick and the spatial extent measured in the B- (or any optical) band is that of the un-attenuated population only. On the other hand, the 5.8$\mu$m probes the dust-reprocessed light from the ongoing starburst and traces the actual spatial distribution of the current star formation. This result confirms that physical sizes of dusty galaxies measured in the UV/optical depend highly on the geometry of the dust distribution, and can be significantly overestimated. Moreover, this result has a direct application to cosmological surveys of dusty, high redshift galaxies, since the size measurements of these sources mostly come from rest-frame HST UV/optical imaging.

\begin{figure}[h] \resizebox{\hsize}{!}{\includegraphics{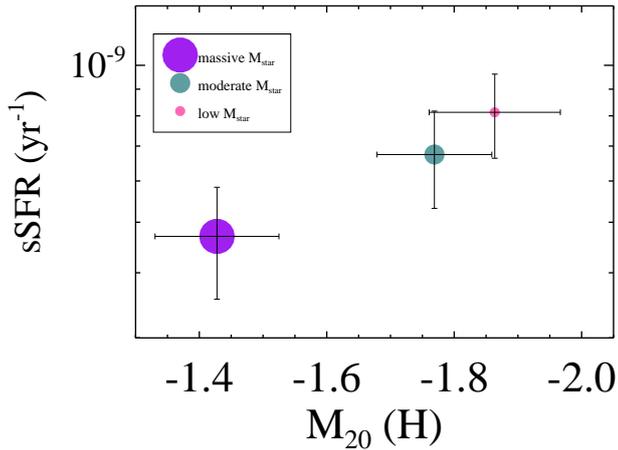}} 
\caption{\small{sSFR-$M_{20}$ in H-band.
The big purple circle represents the mean value of our sample with large stellar masses ($M_{\star}>1.52x10^{11} M_{\sun}$), the intermediate blue circle corresponds to moderate stellar mass ($9.24x10^{10} M_{\sun}<M_{\star}<1.52x10^{11} M_{\sun}$) and the small pink circle show the mean value of small stellar mass ($M_{\star}<9.24x10^{10} M_{\sun}$). The error bars are the standard deviation of the mean values.}}
\label{sSFR_vs_M20_H_H_center_mass_bins_med_err}
\end{figure}

Fig. \ref{sSFR_vs_M20_H_H_center_mass_bins_med_err} shows the sSFR as a function of the $M_{20}$ measured in the H-band for our sample, binned by stellar mass.
We see that more massive galaxies are more extended (i.e., bigger, as you would expect for the same profile, the scaling lengths are larger for the more massive galaxies.). The mean $M_{20}$ values become more negative as the sSFR increases and the mass of the galaxies becomes smaller. Thus, the more massive the LIRG, the more extended the object because the stellar mass is distributed over a larger area. 

%6th project
%=======================  Dust and M20  ================================
\subsection{Dust Temperature and non-parametric coefficients.}
\begin{figure}
	   {\includegraphics[width=\hsize]{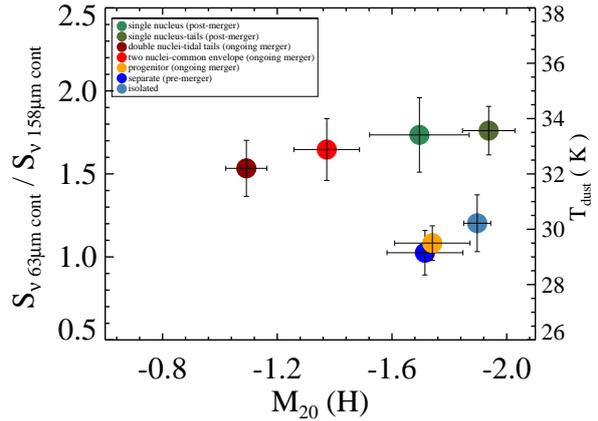}
	   }
\caption{\small{The $M_{20}$ values, calculated in the H-band, along with the corresponding Herschel FIR flux density ratios and the estimated $T_{dust}$. The sample is grouped by the H11 morphology type, also shown in Fig. \ref{Gini_M20_H_center}. The error bars are the standard deviations around the mean of each group.}}
\label{dust_63_158_vs_M20_H_center_seb_class_mean_err_graybody}
\end{figure}

\begin{figure}
	   {\includegraphics[width=\hsize]{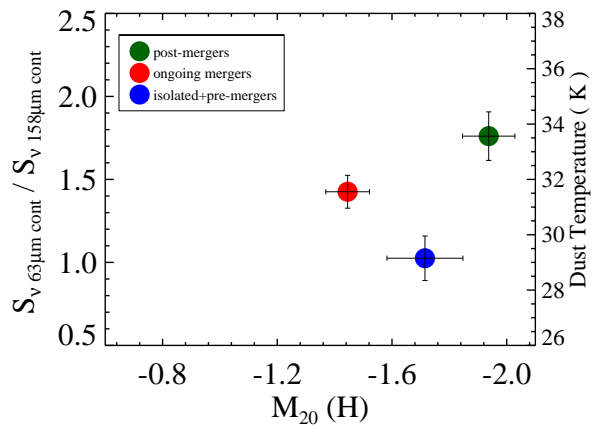}
	   }
\caption{\small{Same as in Fig. \ref{dust_63_158_vs_M20_H_center_seb_class_mean_err_graybody} but now merging the H11 classes to three categories: isolated systems or pre-mergers, ongoing mergers, and post mergers.}}
\label{dust_63_158_vs_M20_H_center_bins_mean_err_graybody}
\end{figure}

We have discussed in the previous sections that LIRGs with larger sSFR appear more compact in the MIR. From \citet{Diaz-Santos10} we also know that the IR compactness of a galaxy is related to its far-IR colors, and therefore to the averaged dust temperature ($T_{dust}$). In this section we explore how $T_{dust}$ evolve along the merger sequence of LIRGs. We can obtain an estimate of the $T_{dust}$ using the $Herschel$ continuum fluxes of 63 and 158 $\mu$m. In particular, we use a modified black body function (gray body) as in \citet{Dupac01} 
\begin{equation}
I(\lambda,T)=B(\lambda,T)(\frac{\lambda}{\lambda_{0}})^{-\beta}, 
\end{equation}\\
where $\beta$=2 and $\lambda_{0}$=100$\mu$m, to fit the Herschel data, and derive the corresponding dust temperature, which is found in the 26-38 K range. \\
\indent In Fig. \ref{dust_63_158_vs_M20_H_center_seb_class_mean_err_graybody} and Fig. \ref{dust_63_158_vs_M20_H_center_bins_mean_err_graybody} we plot the FIR flux density ratio $f_{63\mu m}$ / $f_{158\mu m}$ as a function of the $M_{20}$ values while in the right y-axis we also show the equivalent $T_{dust}$. Pre-mergers and isolated galaxies have low $T_{dust}$ values and more negative $M_{20}$ values. In on-going mergers, the $T_{dust}$ increases drastically and the $M_{20}$ takes higher values (closer to zero), reflecting the separation of the interacting galaxies, in which the starburst has already been triggered. Finally the $T_{dust}$ in post-merger LIRGs increase slightly as the intensity of the starburst event transitions through the peak of star formation. We find that these results are independent of the waveband used to calculate the $M_{20}$ statistic.

%=================================================================================
%================================   DISCUSSION   =================================
%=================================================================================

\section{Discussion} As mentioned earlier number of different studies
including \citet{Lotz04,Hung14,Petty14}  and more recently Larson et al. (2016 in prep.),
 used the $Gini$-$M_{20}$ space in order to describe galaxy morphology based
on a variety of samples of local (U)LIRGs. In general, the analysis used in
these studies is quite similar. They create a segmentation map of every
galaxy excluding the sky pixels and calculate the non-parametric coefficients
inside the map. The major difference between these studies is in the choice of
isophotal threshold level used in the calculation of segmentation map.
\citet{Lotz04} used a value based on the calculation of a
Petrosian-like ellipse. \citet{Petty14} elected to use circular
apertures within the NICMOS FoV, ranging between 7.8$\arcsec$ and
15.4$\arcsec$, for all of their images including the UV and optical while 
\citet{Hung14} defined the pixels of the galaxies according to the method of Quasi-Petrosian
Isophote that \citet{Abraham07} recommended. A novel approach developed by Larson et al. (2016, in prep.) 
uses in turn the surface brightness of galaxies to create more complex segmentation maps that are 
well suited for interacting systems with extended morphological structures, 
and apply the method to a sample of (U)LIRGs in GOALS. For more details, we refer the reader to this work.

For reasons of consistency, in our analysis we classified the
galaxies of our sample using the same area in all three Hubble bands, creating
segmentation maps based on a circular aperture within one Petrosian radius in each band.
As a result, the smaller FoV of the H-band images, combined with the
distance to the individual galaxy of our sample, set an upper limit on
the linear scale and physical extent used to determine its morphology.
We also calculated $Gini$ and $M_{20}$ coefficients following the method
of Quasi-Petrosian Isophote \citep{Abraham07}. We stress here that the
choice to present our results according to the method of one circular
Petrosian because the segmentation maps constructed from Quasi-Petrosian
Isophote method are often contaminated with bright pixels at the edge
the maps.

\citet{Lotz04}, who relied only on R-band imagery and defined the
segmentation maps in a nearly identical manner to our work, showed that
most ULIRGs in the local Universe lie above their defined merger line
and are easily identified by their elevated $Gini$ and $M_{20}$ values.
Our results are not as conclusive. We attribute this to the fact that
GOALS consists of LIRGs of a diverse morphological type and the smaller
FoV which make miss extended emission from clumps or tails away from the
galaxy center. Motivated by \citet{Lotz04}, \citet{Lisker08} concluded
that the $Gini$ coefficient depends strongly on the aperture within
which it is computed and that depends strongly on the depth and quality
of the images. They also suggested that care needs to be taken with the
selection of aperture and limiting magnitude, as well as with the
comparison of calculated $Gini$ values to those of other studies. Our
measurements fully support this conclusion, as is evident by our Fig. 2,
and the discusion in Section 5.1.

Our results are in good agreement with the sub-sample of GOALS
studied by \citet{Petty14}, which also included additional UV imagery.
Quantitatively, our new values of $Gini$, $M_{20}$ are slightly
different, but we also find that they do not effectively reflect the
wavelength dependence of interacting/merging systems, evident in optical
morphology. Merging LIRGs stay in the same general area in the
$Gini$-$M_{20}$ plane independent of wavelength. Despite our lack of UV
data, we also see that as the observed wavelength increases a
significant fraction of ongoing mergers do not substantially change
their locus, even though $M_{20}$ becomes more negative and the $Gini$
values increase. \citet{Petty14} have shown that $Gini$ and $M_{20}$ are
useful in identifying merging LIRGs regardless of rest-frame wavelength
at z $\sim$ 0. Our bigger sample reveals that $M_{20}$ has a larger
dynamic range than $Gini$ and therefore it is more effective in
separating LIRGs at different merger stages, in particular in the
H-band. \citet{Hung14} also studied the merger fraction in a sub-sample
of GOALS and they found that the level of consistency between
$Gini$-$M_{20}$ space and a visual classification was 68\%. They also
stress that LIRGs with disturbed morphology that still have a relatively
smooth light distribution (e.g. advanced mergers with no obvious double
nuclei) often display low $Gini$ values and tend to be classified as
non-interacting systems in the $Gini$-$M_{20}$ plane. Our measurements, presented 
in Section 5.2 are in agreement with this finding.

For the analysis of higher-z sources, \citet{Conselice08} measured
galaxy structure and merger fractions in the Hubble Ultra Deep Field
(HUDF) adopting a circular aperture of 1.5 Petrosian radius. The $Gini$
values were strongly affected by S/N effects for the majority of their
sample. This was also shown in the simulated images of high-z systems
using GOALS galaxies by \citet{Petty14}. The influence low flux pixels
in the calculation of $Gini$ is important because as more low pixels
values enter the segmentation map, the weight of the light distribution
becomes more skewed towards the central high pixel value regions. Along 
the same lines \citet{Kartaltepe10} examined the morphological properties of
a large sample of 1503 70 $\mu$m selected galaxies in the COSMOS field
and they suggest that at z < 1 major mergers contribute significantly to
the LIRG population (from 25 to 40\%) and clearly dominate for the ULIRG
population (from 50 to 80\%). A comparison of their visual
classification to several automated classification techniques commonly
used in the literature (including $Gini$ and $M_{20}$) shows that visual
classification is still the most robust method for identifying merger
signatures because none of the automated techniques is sensitive to
major mergers at all phases. More recently, \citet{Cibinel15} analyzed
Hubble Ultra Deep Field observations, as well as simulated high-z systems, 
and showed that H-band observations alone are not sufficient to trace 
the morphology/structure of stellar masses at high-z, most probably due 
to the fact that they correspond to rest-frame optical emission, also confirming
that their effectiveness can be strongly affected by low S/N. Moreover, 
they demonstrated that adding an asymmetry index to the $M_{20}$ parameter, 
and measuring them in a mass map, rather than an observed near-IR image
can identify mergers with less than 20\% contamination from clumpy disks.

%=================================================================================
%================================   CONCLUSIONS   =================================
%=================================================================================
\section{Conclusions}

In this paper we quantify the galaxy morphology of a sample of 89 LIRGs
from the GOALS sample observed in the optical, NIR, and MIR, using the
non-parametric coefficients $Gini$ and $M_{20}$, we compare their
derived morphology to the one obtained visualy, and explore the consistency 
of the method as a function of a number of physical paramenters. \\
\indent Clearly, our results from the present analysis indicate that simple identification 
of regions in the $Gini$-M$_{20}$ plane as a method to morphologically characterise 
LIRGs in rather challenging. Several parameters related to the rest wavelength 
emission sampled, depth of the imagery along with the size of the FOV can easily lead to their misclassification. 
Revisions to the above methodology, possibly along the lines explored by Larson et al (2016 in prep.) may provide a more robust approach in the future.\\
\indent Our analysis suggests that:

   \begin{enumerate}

    \item The $Gini$ and $M_{20}$ increase in absolute value, 
    when the radius used to create the corresponding segmentation map
    increases. The influence in the calculation of $Gini$ is
    important because as more pixels with low values enter the
    segmentation map, the weight of the light distribution is influenced more
    by the brighter the central regions.

    \item Comparing the B, and I to NIR morphology, we find that the
     median values of $Gini$ increase while median values of $M_{20}$
     become more negative as the wavelength increases.

      \item $M_{20}$ is a better morphological tracer than $Gini$, as it
      can distinguish better systems formed by multiple galaxies
      from isolated and post-merger LIRGs and its effectiveness
      increases with increasing wavelength. In fact, our multi-wavelength analysis allows
      us to identify a region in the $Gini$-$M_{20}$ parameter space
      where ongoing mergers live, regardless of the band used to
      calculate the coefficients. 
      
      \item  We confirm that in B-band, sampling mostly younger stellar
     populations, as the luminosity of the galaxies increases they
     appear more extended and their sSFR increases. In contrast, in
     MIR, (U)LIRGs are more compact than sub-LIRGs. Moreover, the
     sSFR is positively correlated with the $M_{20}$ measured in the mid-IR
     - starbursting galaxies appear more compact than normal ones - and
     it is anti-correlated with it if measured in the B-band. We
     interpret this as evidence of the spatial decoupling between obscured and
     un-obscured star formation, whereby the ultraviolet/optical size
     of LIRGs suffering an intense central starburst is overestimated
     due to the higher dust obscuration towards the central regions.
     
      \item The parameters derived from the $5.8$ $\mu$m image are not
      constraining well the morphology as our
      sample is grouped into unresolved sources concentrated at a given locus
      of the $Gini$-$M_{20}$ plane, while the rest are scattered towards
      higher $Gini$ and lower $M_{20}$ values.

      \item The estimated temperature of the dust $T_{dust}$ increases
      nearly monotonically with the merger state of the galaxies, while
      the $M_{20}$ has a more diverse behaviour from isolated galaxies
      and pre-merger systems (which exhibit more negative $M_{20}$
      values) to on-going mergers (extended objects) and post mergers
      (more compact) regardless of the band.

    \end{enumerate}

%%%%%%%%%%%%%%%%%% Acknowledgements %%%%%%%%%%%%%%%%%
%%%%%%%%%%%%%%%%%%%%%%%%%%%%%%%%%%%%%%%%%%%%%
\begin{acknowledgements}
We thank the referee for the useful comments to the manuscript. We would also like to thank M. Vika (National Observatory of Athens), E. Vardoulaki (Argelander-Institut fur Astronomie) and K. Larson (Caltech) for many useful discussions and comments on this work.
\end{acknowledgements}

%%=================================================================================

\bibliographystyle{aa}
\bibliography{psychogyios_et_al_2016.bib}

%=================================================================================
\begin{appendix} 
%================================== Appendix A ===============================================
\section{The $Gini$ and $M_{20}$ values of the sample in B- and I- field}
\label{appa}
As we discussed in Section 4.1, here we present the $Gini$ and  M$_{20}$ values of our sample in the B- and I-band using the original, uncropped ACS maps in their full 0,05$''$ per pixel resolution. We construct the segmentation map of each galaxy using the same methodology presented in section 4.1. The only difference is that we defined circular apertures using as center the brightest pixel in the I-band image (not the H-band image), and calculate the associated Petrosian radius in both bands. The derived values are present in Table \ref{gini_m20_table_b_i_field}.
\onecolumn

\setlength{\textwidth}{6.8in}
\LTcapwidth=\textwidth
%\scriptsize

\begin{longtable}{lcccc}
\hline 
\hline 
\caption[]{$Gini$, $M_{20}$ values of LIRGs in the B- and I-band using the whole ACS FoV.} \\
\hline  % \\[-2.0ex]
 Optical ID & $ Gini$ (B) & $M_{20}$ (B) & $Gini$ (I) & $M_{20}$ (I)  \\
 (1)&(2)&(3)&(4)&(5) \\
\hline 
\noalign{\smallskip}
\endfirsthead
\hline
\hline
\noalign{\smallskip}
\caption{continued.}\\
\hline 
\noalign{\smallskip}
Optical iD & $ Gini$ (B) & $M_{20}$ (B) & $ Gini$ (I) & $M_{20}$ (I)  \\
(1)&(2)&(3)&(4)&(5) \\
\hline 
\noalign{\smallskip}
\endhead
\hline
\endfoot
\hline
\noalign{\smallskip}
\endlastfoot
\smallskip
  NGC0034                &0.41&-1.21&0.48&-1.77\\
  ARP256N                &0.50 &-1.00 &0.46&-1.37\\
  ARP256S                &0.37&-0.88&0.36&-1.28\\
  MCG+12-02-001          &0.54&-2.56&0.54&-1.21\\
  IC-1623                &0.50 &-1.01&0.44&-0.85\\
  MCG-03-04-014          &0.38&-1.17&0.37&-1.71\\
  CGCG436-030            &0.34&-1.17&0.35&-1.43\\
  IRASF01364-1042        &0.32&-1.16&0.31&-1.52\\
  IIIZw035               &0.49&-1.60 &0.47&-2.03\\
  NGC0695                &0.37&-1.08&0.34&-1.39\\
  PGC9071                &0.42&-1.07&0.39&-1.74\\
  PGC9074                &0.44&-1.32&0.45&-2.23\\
  UGC02369S              &0.53&-0.95&0.49&-0.81\\
  IRASF03359+1523        &0.55&-1.29&0.53&-0.74\\
  ESO550-IG02            &0.43&-1.03&0.49&-0.74\\
  NGC1614                &0.50 &-1.24&0.37&-1.90 \\
  ESO203-IG001           &0.57&-0.73&0.53&-1.00 \\
  VII-Zw-031             &0.37&-1.13&0.35&-1.60 \\
  ESO255-IG007N          &0.31&-0.45&0.44&-0.73\\
  ESO255-IG007S          &0.41&-0.90 &0.41&-1.10 \\
  AM0702-601N            &0.34&-1.36&0.39&-2.11\\
  AM0702-601S            &0.55&-1.55&0.48&-1.41\\
  2MASX-J07273754-0254540&0.49&-0.79&0.37&-0.66\\
  IRAS08355-4944         &0.61&-1.56&0.45&-1.29\\
  NGC2623                &0.38&-1.19&0.41&-1.60 \\
  ESO060-IG016           &0.37&-1.19&0.48&-0.96\\
  IRASF08572+3915        &0.39&-1.17&0.39&-0.93\\
  2MASX-J09133888-1019196&0.39&-1.25&0.46&-0.67\\
  UGC04881               &0.38&-0.81&0.39&-0.94\\
  UGC05101               &0.37&-1.63&0.41&-1.94\\
  IRASF10173+0828        &0.52&-1.95&0.47&-2.12\\
  NGC3256                &0.56&-1.20 &0.48&-1.30 \\
  IRASF10565+2448        &0.56&-0.64&0.53&-0.67\\
  ARP-148                &0.49&-1.08&0.42&-0.98\\
  IRASF11231+1456        &0.40 &-1.36&0.41&-1.97\\
  NGC3690W               &0.50 &-0.97&0.43&-0.88\\
  NGC3690E               &0.50&-0.97&0.43&-0.88\\
  IRASF12112+0305        &0.40 &-0.93&0.43&-1.06\\
  WKK0787                &0.49&-1.59&0.44&-1.75\\
  VV283                  &0.40 &-1.51&0.50 &-2.04\\
  ESO507-G070            &0.39&-1.43&0.44&-1.66\\
  UGC08335W              &0.46&-0.82&0.38&-0.91\\
  UGC08335E              &0.53&-1.98&0.43&-1.85\\
  UGC08387               &0.42&-1.25&0.46&-1.38\\
  NGC5256                &0.39&-0.87&0.39&-0.82\\
  NGC5257                &0.34&-0.64&0.24&-0.79\\
  NGC5258                &0.46&-1.11&0.45&-1.41\\
  UGC08696               &0.36&-1.23&0.36&-1.62\\
  NGC5331S               &0.38&-0.77&0.38&-0.69\\
  NGC5331                &0.43&-1.66&0.44&-1.86\\
  IRASF14348-1447        &0.50 &-1.15&0.50 &-1.27\\
  IRASF14378-3651        &0.39&-1.53&0.42&-2.04\\
  UGC09618S              &0.40 &-1.35&0.41&-1.76\\
  VV705                  &0.42&-1.36&0.46&-1.34\\
  IRASF15250+3608        &0.41&-1.43&0.37&-1.43\\
  UGC09913               &0.28&-1.03&0.34&-1.38\\
  NGC6090                &0.56&-0.90 &0.53&-1.04\\
  2MASXJ16191179-0754026 &0.42&-1.38&0.42&-1.04\\
  ESO069-IG006N          &0.40 &-1.69&0.47&-1.61\\
  ESO069-IG006S          &0.42&-1.82&0.5 &-2.10 \\
  IRASF16399-0937        &0.41&-0.89&0.34&-0.94\\
  NGC6240                &0.30 &-0.96&0.35&-2.13\\
  IRASF17132+5313        &0.43&-0.94&0.47&-0.74\\
  IRASF17138-1017        &0.39&-1.13&0.37&-1.51\\
  IRASF17207-0014        &0.32&-1.07&0.32&-1.35\\
  IRAS18090+0130         &0.38&-0.97&0.41&-1.49\\
  IC4689S                &0.41&-1.12&0.39&-1.45\\
  NGC6670B               &0.48&-0.62&0.58&-0.72\\
  NGC6670A               &0.48&-0.62&0.58&-0.72\\
  NGC6786S               &0.45&-1.43&0.40 &-1.72\\
  ESO593-IG008           &0.40 &-0.86&0.38&-1.14\\
  IRASF19297-0406        &0.45&-1.21&0.56&-1.55\\
  IRAS19542+1110         &0.48&-1.67&0.45&-1.59\\
  IRAS20351+2521         &0.39&-0.98&0.37&-1.39\\
  IIZW096S               &0.41&-1.16&0.51&-0.87\\
  ESO286-IG019           &0.50 &-1.17&0.44&-1.50 \\
  IRAS21101+5810         &0.49&-0.78&0.41&-0.60 \\
  ESO239-IG002           &0.47&-2.18&0.51&-2.44\\
  IRASF22491-1808        &0.38&-0.98&0.55&-1.70 \\
  ESO148-IG002           &0.34&-0.68&0.39&-0.78\\
  IC5298                 &0.42&-1.78&0.45&-2.20 \\
  ESO077-IG014           &0.49&-0.87&0.55&-0.95\\
  NGC7674                &0.45&-0.64&0.49&-0.60 \\
  IRASF23365+3604        &0.31&-1.15&0.36&-1.50 \\
  IRAS23436+5257         &0.40 &-1.17&0.41&-0.98\\
  UGC12812               &0.29&-0.66&0.52&-1.37\\

\label{gini_m20_table_b_i_field}
\end{longtable}
\tablefoot{Columns: (1) Optical cross-identification, where available from NED (see \citet{Armus09} for details). (2),(3) $Gini$ and $M_{20}$ values calculated in the B-band using one Petrosian radius. (4), (5), $Gini$ and $M_{20}$ values calculated in the I-band using one Petrosian radius. We use the brightest pixel in the I-band for the calculation of the Petrosian radius for each galaxy. The Petrosian radius for the ESO203-IG001, ESO593-IG008 and NGC2623 is calculated with respect to the pixel which is close to the bulk of the system. There are three galaxies (ESO099-G004, IRAS18293-3413 and WKK2031) for which we were not able to calculate non-parametric coefficients due to the existence of a large number of field stars surrounding each galaxy.}
\twocolumn

%================================== Appendix B ===============================================
\section{$Gini$ and $M_{50}$}
\label{appb}
\begin{figure}[h] \resizebox{\hsize}{!}{\includegraphics{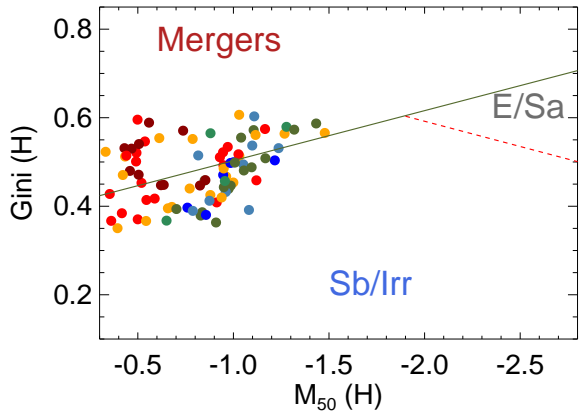}} 
\caption{\small{$G$-$M_{50}$ plane of our sample in H-band. The colored filled circles indicate the different morphological classification of LIRGs as described in Figure \ref{Gini_M20_H_center}. }}
\label{Gini_H_M50_H_H_center}
\end{figure}

As we discussed in the text, the optical and NIR morphologies of LIRGs span the full range from highly disturbed systems to normal spirals, often having fairly bright star forming regions at a large distance   from the central nucleus.  The value of M$_{20}$ is sensitive in tracing bright regions at the outer parts of the galaxies. For that reason, we wanted to examine if a similar non parametric coefficient, the  M$_{50}$  could reveal even better these bright regions. We defined M$_{50}$  according to equation 3 where the limit of $\sigma$M$_i$ is equal to the 50\% of the total flux. In Figure \ref{Gini_H_M50_H_H_center} we present our calculations of $G$-$M_{50}$ in the H-band following the same notation to the one we used in Figure \ref{Gini_M20_H_center}. \\
Comparing these two figures it is clear that M$_{50}$ is not as sensitive since the range of values it takes is smaller. 
\end{appendix}

%%=================================================================================

\end{document}